\documentclass{aa}

\usepackage{graphicx}
\usepackage{txfonts}
\usepackage{longtable}
\usepackage{hyperref}
\usepackage{cancel}
\usepackage{url}
\usepackage[normalem]{ulem}
\usepackage{xcolor}
\usepackage{xspace}

\graphicspath{ {./images/} }

\hypersetup{
    colorlinks=true,
    linkcolor=blue,
    filecolor=magenta,
    urlcolor=cyan,
}

\newif\ifdebug

\debugtrue

\usepackage[utf8]{inputenc}
\usepackage{color}

\newcommand{\gr}{$\gamma$-ray}
\newcommand{\odaurl}{https://www.astro.unige.ch/cdci/astrooda_}

\newcommand{\change}[1] {#1}

\newcommand{\integral}{\textit{INTEGRAL}\xspace}
\newcommand{\xmm}{\textit{XMM-Newton}\xspace}
\newcommand{\swift}{\textit{Swift}\xspace}
\newcommand{\fermi}{\textit{Fermi}\xspace}

\newcommand{\isgri}{\textrm{ISGRI}\xspace}
\newcommand{\ibis}{\textrm{IBIS}\xspace}
\newcommand{\jemx}{\textrm{JEM-X}\xspace}

\begin{document}

\title{An online data analysis system of \integral telescope}
   \author{A. Neronov,
          \inst{1,2}
          \and
          V. Savchenko\inst{1}
          \and
          A. Tramacere\inst{1}
          \and
          M. Meharga\inst{1}
          \and
          C. Ferrigno\inst{1}
          \and
          S. Paltani\inst{1}
          }

   \institute{
             {Department of Astronomy, University of Geneva, Ch. d'\`Ecogia 16, 1290, Versoix, Switzerland}
         \and
             {APC, University of Paris, CNRS/IN2P3, CEA/IRFU,  10 rue Alice Domon et Leonie Duquet, Paris, France}
             }
\date{}

\abstract
   {During more than 17 years of operation in space \integral telescope has accumulated large data set that contains records of hard X-ray and soft \gr\ astronomical sources. These data can be re-used in the context of multi-wavelength or multi-messenger studies of astronomical sources and have to be preserved on long time scale{s}.}
   {We present a scientific validation of an interactive online \integral data analysis system for multi-wavelength studies of hard X-ray and soft \gr\ sources.  }
   {The online data analysis system generates publication-quality high-level data products: sky images, spectra and light-curves in response to user queries that define analysis parameters, such as source position, time and energy interval and binning. The data products can be requested via a web browser interface or via Application Programming Interface (API) available as a Python package. The products for {the} \ibis/\isgri instrument of \integral are generated using the Offline Science Analysis (OSA)  software which is provided by the instrument teams and is conventionally used to analyse \integral data. The analysis workflow organized to preserve and re-use various intermediate analysis products, ensuring that frequently requested results are available without delay. The platform is implemented in a Docker cluster which allows operation of the software in a controlled virtual environment, and can be deployed in any compatible infrastructure. The scientific results produces by ODA are identical to those produced by OSA, since ODA simply provides a platform to retrieve the OSA results online, while leveraging a provenance-indexed database of pre-computed (cached) results to optimize and reuse the result.}
   {We report the functionalities and performance of the online data analysis system by reproducing the benchmark \integral results on different types of sources, including bright steady and transient Galactic sources, and bright and weak variable  extra-galactic sources. We compare the results obtained with the online data analysis system with previously published results on these sources. We also discuss limitations of the online analysis system.}
   {We consider the \integral online data analysis as a demonstrator of more general  web-based ``data analysis as a service'' approach that provides a promising solution for preservation and maintenance of data analysis tools of astronomical telescopes  on (multi)decade long time scales and facilitates combination of data in multi-wavelength and multi-messenger studies of astronomical sources. }


   \maketitle
\section{Introduction}

The INTErnational Gamma-RAy Laboratory (\integral)  \citep{integral} is an ESA space astronomy mission {that has collected} data in orbit since 2002. It provides {observations}  of astronomical sources in the keV-MeV energy range. {IBIS, the Imager on Board the \integral Satellite \citep{ibis}, is a coded-aperture instrument that provides fine imaging (12$^\prime$ FWHM) in a nearly squared partially coded field-of-view (FoV) of 30$^\circ\times$30$^\circ$, source identification and spectral sensitivity to both continuum and broad lines between 15 keV and 10 MeV. Its focal plane is composed of two detectors, optimized for two different energy ranges: \isgri from 15 to $\sim$1000 keV \citep{isgri} and PICsIT from 400 keV to 10 MeV \citep{picsit}.} In the following, we will discuss only the \isgri detector layer, and refer to it as the \isgri instrument. The Joint European X-Ray Monitor \citep[JEM-X][]{jemx} provides images, timing, and spectral information in a lower energy range (3--25 keV). It has a narrower circular FoV $\sim 10^\circ$ diameter. The spectrometer SPI \citep{spi} provides high-resolution spectroscopy data with moderate angular resolution. The SPI instrument is surrounded by an Anti-Coincidence Shield (ACS) that reduces the level of background in SPI and simultaneously works as an all-sky Gamma-Ray Burst (GRB) monitor \citep{spiacs}.
The wide FoV of IBIS inherently includes data on a large number of astronomical sources, which are not necessarily the targets of specific observational proposals. This means that all source-specific observational campaigns of \integral possess large potential for serendipitous science. All \integral data become publicly available after a one-year proprietary period, so that analysis of all the sources found in the field of view is possible.

Astronomical sources visible in X-ray and soft \gr\ band are variable on different time scales, from milliseconds to years and decades \cite{variabilityXray,variabilityXray_agn,multimessenger}. In this respect, the 17-year-long data-set of \integral is unique, as it contains the information on long-term variability patterns of a large number of sources. Some sources spend long periods in quiescent states producing little emission and exhibit occasional outbursts or flares on irregular basis with a duty cycles spread over years or decades. The \integral archive contains data taken nearly continuously since min November 2002 with a duty cycle on nearly 90\% -instruments are switched off at each perigee passage through the Earth radiation belts. Until March 2003, on-board setting were continuously adjusted to optimize the performance, making the scientific analysis a challenging. The general user is normally suggested to start using archive data after this date finding nearly 50\,Ms of usable data in the direction of the Galactic center, the region with the highest exposure. Archive data contain, thus, information on previous history of quiescence and activity of sky sources ($>1200$ detected by the imager so far), including those which might undergo outbursts in future.

The \integral data and Offline Science Analysis (OSA) software that includes data analysis pipelines for {all instruments, including} \isgri \citep{isgri-da} and JEM-X \citep{jemx-da}
is distributed by the \integral Science Data Centre \citep[ISDC,][]{isdc}.
The main components of OSA have been developed in the period preceding the \integral launch and are maintained by the ISDC. The architecture of OSA was optimized for computing environments that were common more than 20 years ago. \integral was initially planned to operate in space for five years and generate relatively small data sets. The only solution for data processing was via local installation of OSA on a user computer. This is not necessarily the case today for data sets spanning 17 years. Their analysis requires a significant amount of computing resources. Moreover, maintenance of legacy software on evolving operating systems poses more and more challenges.

The development of high-performance computing (HPC) and cloud computing (CC) technologies and their applications to astronomical data management \citep{lsst_kubernetes,astronomy_clouds} over the last decades opens up a new possibility of deployment of OSA-based data analyses without the need for local installation of the software, and provides access to large pool of computing resources, thus significantly reducing the complexity of the analysis of \integral data. Such online data analysis (ODA) system for the \isgri instrument has been recently developed at the Department of astronomy of the University of Geneva\footnote{\url{https://www.astro.unige.ch}} in synergy with the \integral Science Data Centre (ISDC)\footnote{\url{https://www.isdc.unige.ch/integral/}}, and is maintained jointly with the François Arago Centre (FACe) of Astroparticle and Cosmology laboratory in Paris\footnote{\url{http://www.apc.univ-paris7.fr/FACe/home}}.

This system entirely relies on the official OSA package, provided and validate by the instrument teams, integrated and distributed by by ISDC - all of the results are not only equivalent, they are identical. This design allows to leverage, in principle, the unrestrained power of the OSA software, preserving and maintaining the complete potential of the \integral data for the future explorers, without making assumptions about which products are likely to be useful. While this approach may require larger real-time computing resources, the platform exploits a dynamic innovative provenance-based database of precomputed products to eliminate analysis duplication and flexibly respond to the user requests, as detailed in Section~\ref{sec:storage} and Section~\ref{sec:odavsworld}. This means, in principle, that the ODA validation is in part redundant, since it is equivalent to OSA validation. On the other hand, this paper reveals how a selection of particular OSA cookbook threads, adopted in ODA, allows to reproduce the published results.

In the following, we describe the system and characterize its performance via comparison of science results obtained using the system with previously published benchmark \integral results on a variety of Galactic and extra-galactic sources. We also demonstrate the usefulness of the system in the context of multi-wavelength studies and discuss {advantages and} limitations of {this online data analysis implementation}.

{Throughout the paper, we  provide direct links to the code producing the figures in the text, showing the potential of this approach for the open access, reusability and reproducibility of science results.}

\section{Online data analysis interface}
\begin{figure}
\includegraphics[width=\linewidth]{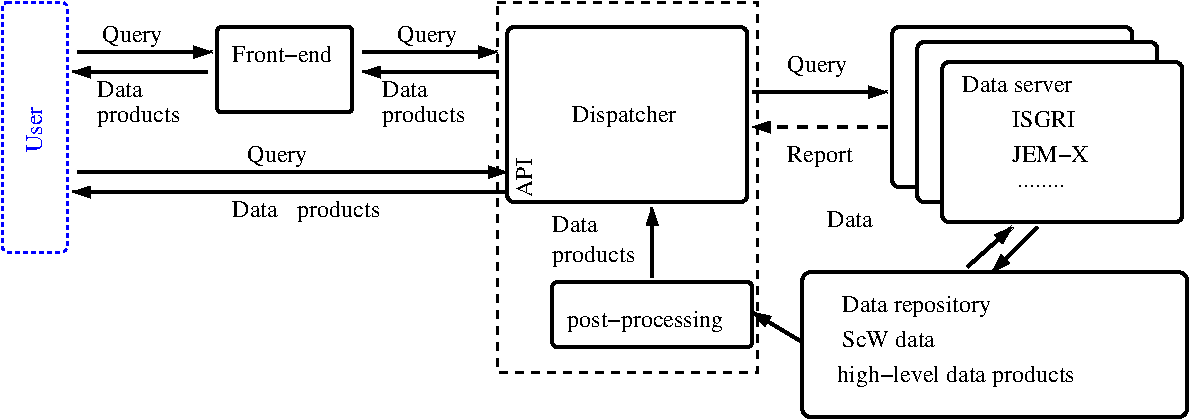}
\caption{Architecture of the \integral Online Data Analysis system.}
\label{fig:scheme}
\end{figure}

The general scheme of the ODA is shown in Fig. \ref{fig:scheme}. The user has several possibilities to submit requests for analysis of \integral data:
\begin{itemize}
    \item through a web browser by accessing the ODA website\footnote{\url{https://www.astro.unige.ch/cdci/astrooda_}} on his/her local computer  and entering the analysis parameters into the parameter boxes, or
    \item directly specifying analysis parameters in a URL link to the ODA website (examples are given in the next section) or
    \item through an ODA Application Programming Interface (API), {\tt oda\_api}\footnote{\url{https://github.com/oda-hub/oda_api}}, e.g., from a Jupyter Notebook, also on their local computer.
\end{itemize}

The full process can be schematized as follows:
\begin{enumerate}

\item  The request can be sent using the front-end or the {\tt oda\_api} client.  Both interfaces verify the syntactical  correctness and completeness of user queries.

\item  Requests arrive at dispatcher and here processed by  an internal abstraction process, which implements classes (interfaces) for specific instruments data products, such as spectra or light curves data, and post-processing products, such as mosaic images, light curves  images, spectral fits.

\item Each data  product interface communicate with a specific backend, i.e. data server(s) implemented as Docker\footnote{\url{https://hub.docker.com/r/odahub/ddosa-interface}} containers running OSA in service mode. The containers are currently deployed locally on the HPC resources of the University of Geneva, but they could be deployed on any other HPC or CC services.

\item The request can be either synchronous or asynchronous. In the latter case, a continuous report from the backend brings information to the dispatcher regarding the process status until the product is ready. In the former case, a single report is provided.

\item Data products provided by the data server upon analysis requests are stored in a data repository and are made available to the dispatcher.

\item In the current version of ODA, the dispatcher also performs post-processing and visualisation of the data, using specific services, providing high-level products to be displayed on the front-end.

\item Final products are available to the user either through the front-end or through the client API. The front-end displays sky images, spectra, light curves, and source catalogs in a virtual-desktop environment embedded in the web browser, providing the possibility to download the data products to the local user computer.
\end{enumerate}

\subsection{Front-end interface -- Dispatcher -- Data Server interactions}

\begin{figure*}
\includegraphics[width=0.78\linewidth]{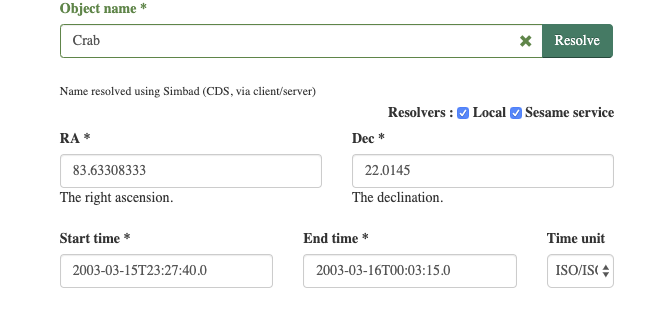}
\includegraphics[width=0.2\linewidth]{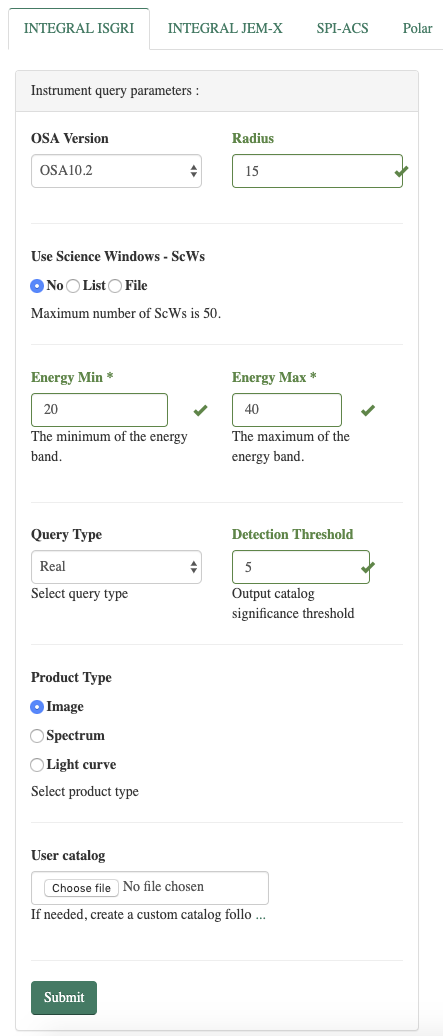}

\caption{Left: General parameter space of ODA front-end. Right: example of instrument-specific parameter field for \isgri telescope. }
\label{fig:general_parameters}
\end{figure*}

{\ibis and \jemx are coded-mask instruments that rely on a dithering pointing strategy with individual exposures called Science Windows (ScW) lasting 0.1--1 hour. Reconstruction of sky images creates a catalog of detected sources. This catalog should be used during the extraction of spectra and light curves of specific sources. Disentangling of signals from different sources in the field of view requires a sky model which lists positions of all possible sources. \citep[see][for details]{isgri-da}.
The default ODA workflow allows the user to select the data set, obtain a catalog of detected sources by reconstructing an image, and then manipulate this
catalog to extract spectra and light curves.}

The data processing  is initiated in response to the user queries defining the analysis parameters (Fig. \ref{fig:general_parameters}). These queries include at the very minimum:
\begin{itemize}
\item Source name or sky coordinates
\item Time interval for the data or Science Window (ScW) list
\end{itemize}

The front-end is able to resolve astronomical source names by accessing two of the main astronomical databases (SIMBAD and NED). It accepts time parameters in several different conventional formats. The list of ScWs can be specified as a comma-separated list of their unique identifiers. The ScW data base is separately accessible through the w3browse interface on the web pages\footnote{\url{https://www.isdc.unige.ch/integral/archive}}.

Apart from these generic query parameters, the front-end allows the user to specify  parameters that are  specific to the \integral instruments: \isgri, JEM-X, SPI-ACS. An example of parameter field for \isgri is shown in the right panel of  Fig. \ref{fig:general_parameters}. For \isgri and JEM-X, it is possible to specify:
\begin{itemize}
\item one of the two currently used versions of OSA: 10.2 and 11.0;
\item radius of the ``region of interest'' within which pointing data are selected (which depends on the instrument field-of-view);
\item One of the two {units of the JEM-X instrument} (JEM-X1 or JEM-X2);
\item type of the data product (image, spectrum or light curve);
\item energy range of the image  and light curve. It should be noted that the spectrum is always computed in the full energy range with predefined resolution (16 channels for JEM-X, 128 channels for \isgri;
\item minimal detection significance of sources in the output catalog from imaging;
\item time binning for the light curve;
\item source catalog to be used in spectral and timing analyses.
\end{itemize}

In a similar way, the parameters can also be specified in the API requests using the {\tt oda\_api} Python package. For example, OSA version is specifiable by setting a parameter {\tt osa\_version='OSA10.2'} in the API requests\footnote{See {\tt oda\_api} documentation at \url{https://oda-api.readthedocs.io/en/latest/} for full details.}.

The front-end provides a display of the high-level data products (images, spectra, light curves, source catalogues) through a virtual desktop environment. It also provides the possibility of performing post-analysis of the data, like, e.g., fitting spectra with XSPEC spectral models, using public domain  rendering packages\footnote{\url{https://bokeh.pydata.org/en/latest/}}.

\subsection{Data analysis and storage organization}
\label{sec:storage}
The ODA infrastructure is using the online archive of \integral raw data, provided by the ISDC\footnote{\url{http://www.isdc.unige.ch/integral/archive}}.
The data server's task is to provide high-level data products corresponding to the user requests received through the dispatcher. Running OSA is time consuming (about 50 CPU-hours for a spectrum in a  typical short transient observation, or of the order of 2000 CPU-hours for an analysis of historic data for a typical source), so as far as possible it is desirable to keep pre-computed products for future (re-)uses. However, it is not possible to store high-level data products for all imaginable combinations of user input parameters. In ODA, the permanent archive of the raw data is complemented by an analysis cache containing high- and intermediate-level data products that are added or removed depending on the the user demands averaged over certain time periods.

The cache storage is organized according to the data lineage \cite[e.g.][]{Ikeda2009}, which is a specific case of data provenance \citep{Gupta2009}. The data lineage metadata comprises the information on the sequence of analysis steps (the analysis or workflow nodes) undertaken to produce given high- or intermediate-level data products. The ontology of the workflow nodes prescribes specific metadata associated with each step, and induces the collection of metadata of the final product.

The lineage metadata of the cache storage contains all relevant properties of all stored data products, and only this information. This provides a possibility to re-use previously produced intermediate-stage data products for the processing of new data analysis requests by users. New data analysis workflows typically do not start processing of raw data from ``from scratch''. Instead, they are formed {from a combination of parts of already  available workflows derived from} specific intermediate or high-level data products stored in the cache, together with the provenance DAG (Directed Acyclic Graph) metadata. This approach provides an efficient way to speed-up data processing following user requests if those are repetitive or recursive or if the requests are nearly identical to those done by previous users with only moderate parameter modifications.

Efficient reuse of parts of the OSA based data analysis workflow is enabled by the re-organisation of OSA  in data analysis units expressed as Python classes, following the Declarative Data Analysis (DDA) approach inspired by the principles of functional programming. This development was driven by the needs of efficiently managing the data of \integral together with the information on the 17-year history of the telescope operations: by 2018, in the raw-data archive, there are about $10^3$ different types of data occupying 20 TB in some $2\times 10^4$ files.

The re-factored OSA implementing the DDA approach (called DDOSA\footnote{See \url{https://github.com/volodymyrss/dda-ddosa/} for implementation details.})
follows a simple logical scheme suitable for  reproducibility of the analysis.
Each analysis  unit is a pure function of its input data, meaning that it depends only on its own explicit input. It transforms the input data into other  data products. Any data are uniquely identified by a tree of connected analysis units that were used to produce it, or, equivalently, by its DAG  ``provenance graph''. In other words, DDOSA uses provenance as a data identifier \citep[see][for more details]{Savchenko2019_oda}.

\begin{figure*}
	\centering
 	\includegraphics[width=1.\linewidth]{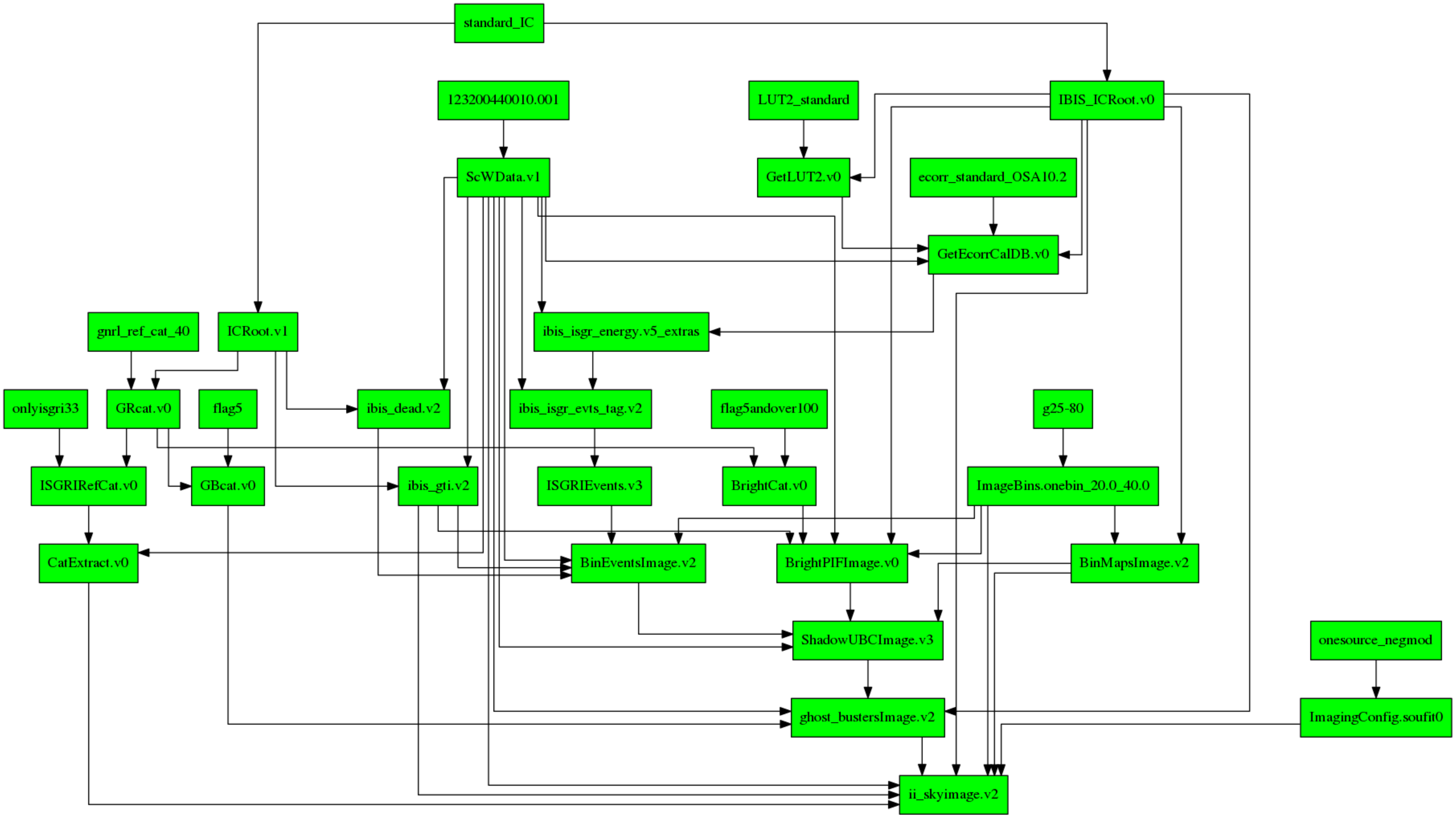}
	\caption{Example of high-level provenance graph for a sky image derived from a single \integral/\isgri ScW.}
\label{fig:transient}
\end{figure*}

The high-level data products associated to very large analysis chains may be eventually associated with a very large provenance graphs. An example of the provenance graph for a single ScW image high-level data product is shown in Fig.~\ref{fig:transient}.

The DAG provenance graph approach for data identification at different analysis levels is optimal not only for caching frequently re-used intermediate analysis step results, but also for the parallelization of the analysis. The DAG structure of DDOSA workflows implies the presence of different independent branches of analysis that can be naturally executed  independently in a distributed environment. This is taken into account in the system of job scheduling. For each analysis unit, execution requests originate either from the users (via dispatcher) or other analysis units. Each request processing starts from the evaluation of the request, resulting either in the retrieval of the requested products from the storage cache or in the delegation of the request to a local or remote  executor, a process which is transparent from the point of view of the request.
A simple scheduling layer has been implemented following this approach. The advantage of this scheduler is the straightforward treatment of complex dependencies.

\section{Benchmark analysis results}

The web-based ODA interface retains all the functionalities of OSA and could be used to obtain publication quality results of analysis of \integral observations {with no difference from what an experienced user can obtain running locally OSA}. In this section, we demonstrate the performance of the ODA based analysis on benchmark cases, by showing that the results obtainable with ODA are compatible with previously published results or improve on them, owing to upgraded algorithms and calibration. 

\subsection{Crab pulsar \change{and nebula}}
\label{sec:crab}

The Crab pulsar and Nebula complex is one of the brightest  sources on the {X-ray} sky \citep{crab_review}. Because of this property {and its flux stability},
it is often used as a ``calibration'' source in high-energy astronomy. \integral observations of Crab are reported in a number of publications
\citep[e.g,][]{crab} and, in particular, in the context of {the} study of {the} long-term variability  of the source {emission} \citep{crab_variability}. We
verify the performance of  ODA by reproducing the published result{s} on Crab variability and extending the evolution study {over} a 15 year time period. It can
be noted that the magnitude and details Crab of variability as observed by \isgri is not identical to those reported by other instruments. The excess variability
in in part due to the systematic uncertainties in \isgri calibration, and in part due to the difference in the instrument energy ranges.  Detailed evaluation of
if OSA reconstruction of \isgri observations as well as discussion \isgri calibration challenges is beyond the scope of this work. Here we demonstrate that ODA
reproduces the best available OSA results. ODA will naturally follow any upgrades to OSA software and calibration files in the future.

\begin{figure*}
\includegraphics[width=\linewidth]{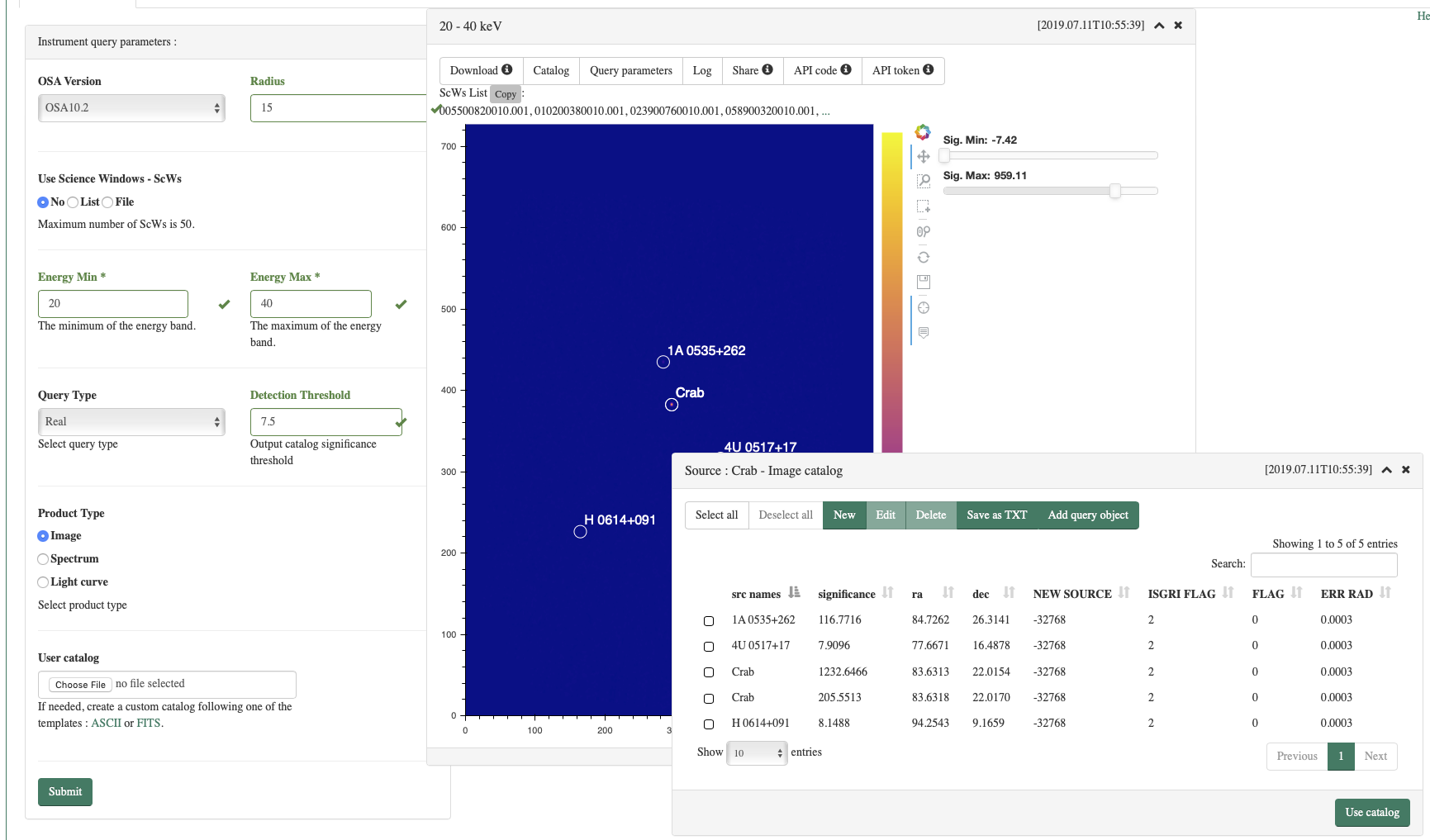}
\caption{Mosaic image of Crab region   extracted from a sample of 50 randomly selected ScWs, together with the display of the catalog of detected sources. The result could be re-generated using the URL \url{https://doi.org/10.5281/zenodo.3634648}.}
\label{fig:Crab_image}
\end{figure*}

\begin{figure}
\includegraphics[width=\linewidth]{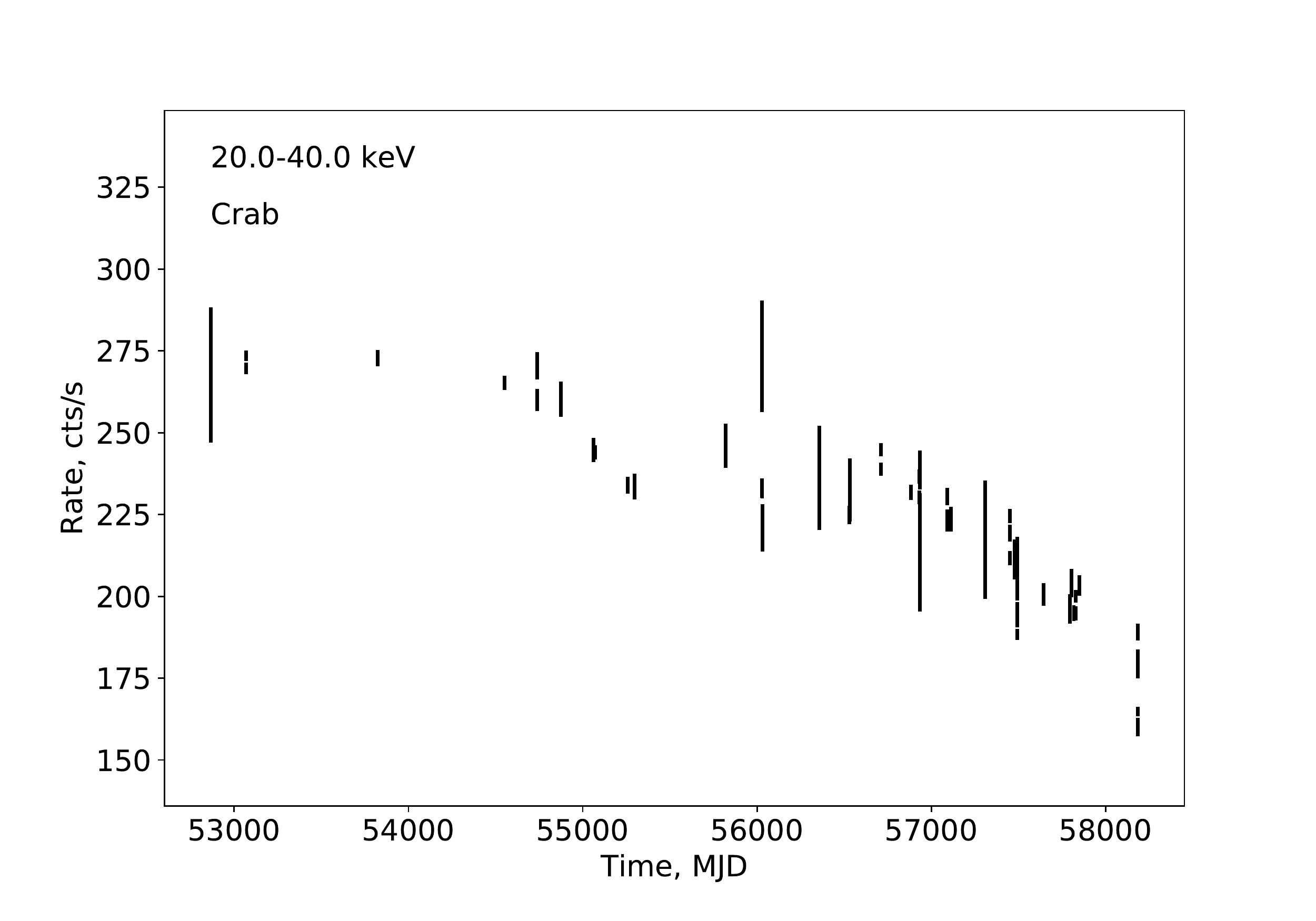}
\caption{Crab long-term (15 year time span) lightcurve extracted from a sample of 50 randomly selected ScWs in 30-100 keV band, using OSA10.2. The lightcurve is
    generated via  \href{https://github.com/oda-hub/oda_api_benchmark/blob/master/examples/Crab_spectra_lc.ipynb}{Crab\_spectra\_lc.ipynb} via
    \href{https://github.com/oda-hub/}{oda-hub}.}
\label{fig:Crab_lc}
\end{figure}

The ODA interface is currently limited to single requests for data analysis based on no more than 50 {science windows (ScWs)}, to limit the waiting time on the available resources (see Section~\ref{sec:limitations} for details, future plans, and a work-around). If the requested time span of data extends over more than 50 ScWs, random selection of ScWs within the specified time limits is performed. This is the case for the results reported in Figs. \ref{fig:Crab_image}, \ref{fig:Crab_spectrum}, and \ref{fig:Crab_lc}. The time interval of the analysis for this specific ScW subset is 2003-03-15 23:27:40.0 to 2018-03-16 00:03:15.0 UTC, spanning over more than 15 years. Pointings within 15 degrees from Crab position are selected.

{Our example Crab image} could be accessed or re-generated by directly re-launching the analysis via an URL \href{https://www.astro.unige.ch/cdci/astrooda_}{https://www.astro.unige.ch/cdci/astrooda\_} in which the analysis parameters are specified after the sign \verb+{?}+ and separated by the \verb+{\&}+ sign as, for example: {\tt\small src\_name=Crab\&RA=83.633080\&DEC=22.014500\&radius=15} for the source name and/or sky coordinates and region of interest specification, {\tt\small  \&T1=2003-03-15T23:27:40.0 \&T2=2018-03-16T00:03:15.0 \&T\_format=isot} for the time interval, {\tt\small  \&E1\_keV=20 \&E2\_keV=40} for the energy interval, {\tt\small  \&instrument=isgri \&osa\_version=OSA10.2 \&product\_type=isgri\_image} for the instrument, software version and data product specification (spaces are added for readability, but should be removed in the actual query). The parameter {\tt\small detection\_threshold=7.5} will result in display of the sources detected at significance level higher than $7.5$. The analysis with specified parameters is launched automatically, as soon as the instrument parameter is defined: {\tt\small  \&instrument=isgri}. The parameters {that} are not explicitly specified in the parameter field of the URL are fixed to their default values.

The  executable URL with all specified parameters for each data product could be obtained by pressing the ``Share'' button displayed in the data product window on ODA frontend (see Fig. \ref{fig:Crab_image}).

{This example} {of} analysis could also be launched from a python interface on the user laptop (e.g., from a {shell or a} Jupyter notebook) by providing parameters to the request
{\small \begin{verbatim}
from oda_api.api import DispatcherAPI
disp=DispatcherAPI(
    host=
    'www.astro.unige.ch/cdci/astrooda/dispatch-data')
disp.get_product(
    RA=83.633080,
    DEC=22.014500,
    radius=15,
    T1='2003-03-15T23:27:40.0',
    T2='2018-03-16T00:03:15.0',
    E1_keV=20.0,
    E2_keV=40.0,
    instrument='isgri',
    product='isgri_image',
    osa_version='OSA10.2')
\end{verbatim}}

The API code for each data product {can} be obtained directly by pressing the ``API code'' button in the product window on {the} ODA front-end (see Fig. \ref{fig:Crab_image}).

A crucial part of the imaging analysis is the search for significantly detected sources both in individual ScWs and in the mosaic image. Setting the source
detection threshold to $7.5\sigma$ (a parameter in the web form of ODA) results, {in our example}, in the detection of four sources displayed in the image in
Fig. \ref{fig:Crab_image}. Details of the source detection are available in the catalog display accessible through a button  from the image display panel, as
shown in Fig.~\ref{fig:Crab_image}. Occasionally, sources may have multiple appearances in the catalog display, because this table combines several output
catalogs of {the standard \integral} analysis, {namely} results of the search of sources in the mosaic and in individual ScWs. This might be important, because
some flaring sources are detectable in individual ScWs during short time periods, but are not detectable in mosaic images {with} longer exposure times (as it is
typical for bursting and transient sources, see e.g..\citealt{magnetar} for a recent example). The user is asked to carefully inspect the output catalog from the
imaging step and adjust the source selection for the following spectral and timing analys{es}.

Imaging, spectral extraction and timing routines of OSA use catalogues of sources to match the shadow patterns corresponding to {these} sources on the detector
plane. The catalog used for imaging, spectral or timing analysis could be explicitly specified in the ``User catalog'' parameter window in the parameter panel.
If no user catalog is specified, the default general \integral catalog is used. {This is advisable for the imaging products, but sub-optimal for the extraction
of spectra and light curves, which {relies} on fitting a sky model on the shadowgram. If this sky model is redundant, the fitting becomes more problematic,
resulting in unreliable flux determinations.} The user {can} edit the catalog entries in the display of the catalog output of the imaging step. This display
also has {a} ``Use catalog'' button{,} which would push the edited catalog to the ``User catalog'' to be used at the subsequent stages of analysis. The catalog {can} also be defined explicitly in the form of a python ``dictionary'' in the URL parameter field. The correctly formatted catalogue embedded {in the} URL
{can} be obtained by clicking the ``Share'' button next to the displayed data product.

The display of results of spectral analysis for all sources listed in the catalog (user custom catalog or the output catalog from the imaging step analysis)
provides a possibility to choose {a} spectral model for fitting the spectrum\footnote{Based on Xspec package fitting
(\url{https://heasarc.gsfc.nasa.gov/xanadu/xspec/}).}. The display of the spectrum together with the fitted model also provides the details of the fit, such as
the model parameter values with their uncertainties. Binning of the spectra is performed only at the plotting stage, the fit is performed on the spectrum at
full resolution, which can be downloaded from the web interface.
\change{We notice that} the 20-30~keV band is  affected by the long-term evolution of the \isgri response, as the \isgri energy threshold is gradually increasing
with time and {low-energy events} are lost. {For consistency, only data above 30 keV are thus automatically fitted in the web interface, but data at lower
energy are available, upon download of the FITS-format spectral file\footnote{\url{https://www.isdc.unige.ch/integral/download/osa/doc/10.2/osa_um_ibis.pdf}}}.

Fig. \ref{fig:Crab_lc} shows the 30--100 keV light-curve of the source during a 17-years time span, extracted from the same set of 50 random ScWs and binned
into ScW-long time bins. The {figure} shows the fit of the light-curve with constant and linearly changing flux models. There is a noticeable decrease of the
source count rate toward later times, which become{s} especially pronounced after MJD 56800 (mid-2014). Superimposed onto this {instrumental} trend, there is
the true variability of the Crab nebula studied by \citet{crab_variability}.

Such rapid decrease of the count rate is due to the decrease of the instrument response at low energy and is not corrected in version 10.2 of OSA, because
calibration algorithms were not able to correct this rapid evolution and
calibration files were frozen at this moment. The correct instrument response after MJD 56800 is provided by version 11.0 of OSA with the relative calibration
files\footnote{{It is foreseen that the OSA11 software release will cover the full mission at the end of 2020.}}. See Section~\ref{sec:limitations} for details
on the instrumental effects contributing to these results.

\begin{figure*}
\includegraphics[width=\linewidth]{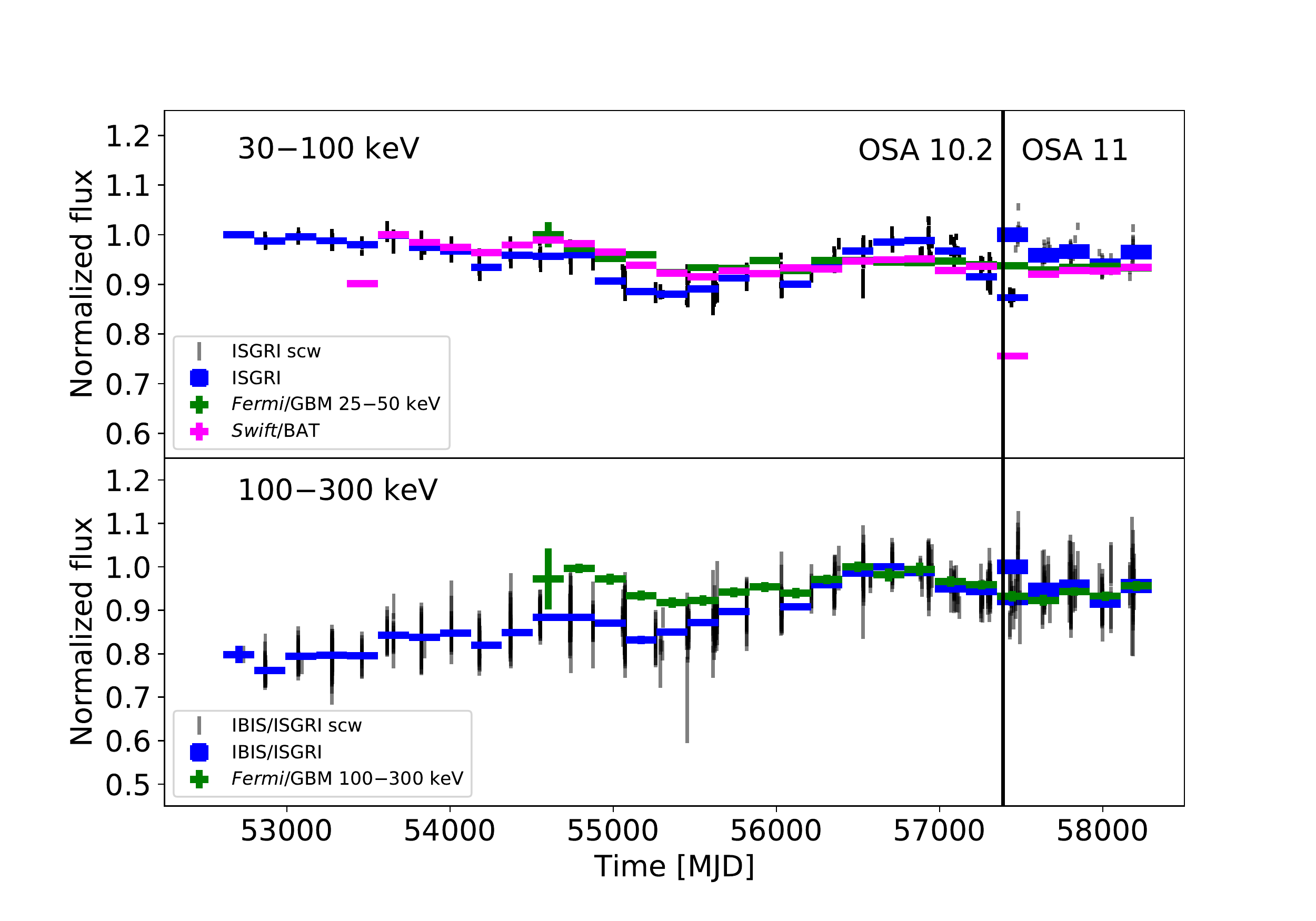}
\caption{Evolution of the Crab count rate over 17 years of \integral operations in 30--100 keV (top panel), 100--300 keV (bottom panel) energy ranges, compared to
    {those} observed by \swift/BAT (magenta) and \fermi/GBM (green data points) \citep{crab_variability}. Grey data points show the lightcurve binned in ScW by
    ScW time bins by the ODA. Blue blue data points show the rebinn{ed} individual ScW measurements. The reader can launch the Jupyter notebook
    \href{https://github.com/oda-hub/oda_api_benchmark/blob/master/examples/Crab_lc_longterm.ipynb}{Crab\_lc\_longterm.ipynb} used to generate these long-term
    light curves via API access to ODA.}
\label{fig:crab_variability}
\end{figure*}

Fig.~\ref{fig:crab_variability} shows a comparison of the long-term variability of the Crab flux in the 30--100 and 100--300 keV ranges for 17 years of \integral
operations together with the ones measured by \swift/BAT and \fermi/GBM telescopes \mbox{\citep{crab_variability}}.
To produce this figure, we select random sets of 50 ScWs spanning one year time intervals and extracts ScW-by-ScW lightcurves by specifying 10\,ks time steps in the ODA parameters for lightcurve time binning (this is longer than the duration of one ScW). These lightcurves are subsequently averaged into the time bins used by \citet{crab_variability} for comparison. For this workflow, we exploited the API access to ODA platform, as coded in the \href{https://github.com/oda-hub/oda_api_benchmark/blob/master/examples/Crab_lc_longterm.ipynb}{Crab\_lc\_longterm.ipynb} Jupyter notebook, which is part of the \href{https://github.com/oda-hub/oda_api_benchmark}{https://github.com/oda-hub/oda\_api\_benchmark}  GitHub repository.
In Sect.~\ref{sec:3c}, we detail how to run online Jupyter notebooks as the one used to produce Fig. \ref{fig:crab_variability}.

From Fig. \ref{fig:crab_variability} one {can} see that the lightcurve extracted using OSA10.2 and OSA11.0 in their time intervals of validity are compatible
with the lightcurves of \swift/BAT and \fermi/GBM, {after they have been normalized to their average level.   Intrinsic variability of the Crab nebula (at $\sim
    5\%$ level) can be appreciated in the general trend of all instruments, as done by \citet{crab_variability}.
There is, however, some residual differences between the \integral and \swift/BAT or \fermi/GBM lightcurves, which can be used to estimate the
systematic cross-calibration uncertainty. This
amounts up to 10\% in the 100-300 keV energy range.} Count rate light curve have variations that are also due to
the evolution of the instrument gain; these are accounted for using response files to recover the intrinsic source flux in the spectral extraction stage.

\begin{figure}
    \centering
    \includegraphics[width=\linewidth]{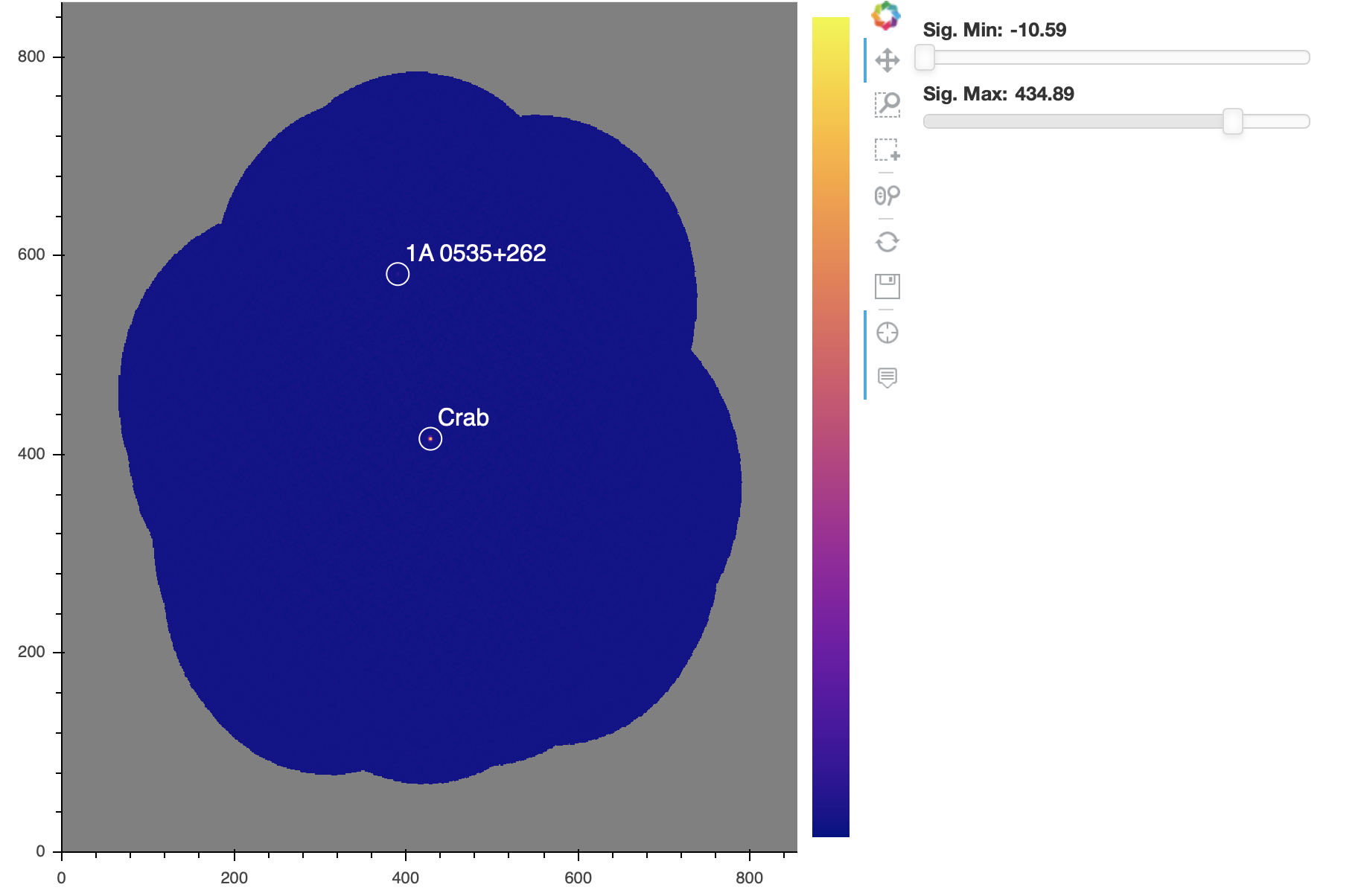}
    \caption{Image of sky region around Crab obtained by JEM-X1 instrument in the energy range 3-20 keV. The image can be regenerated via URL \url{https://doi.org/10.5281/zenodo.3832080}.}
    \label{fig:crab_image_jemx}
\end{figure}

\begin{figure}
    \includegraphics[width=\linewidth]{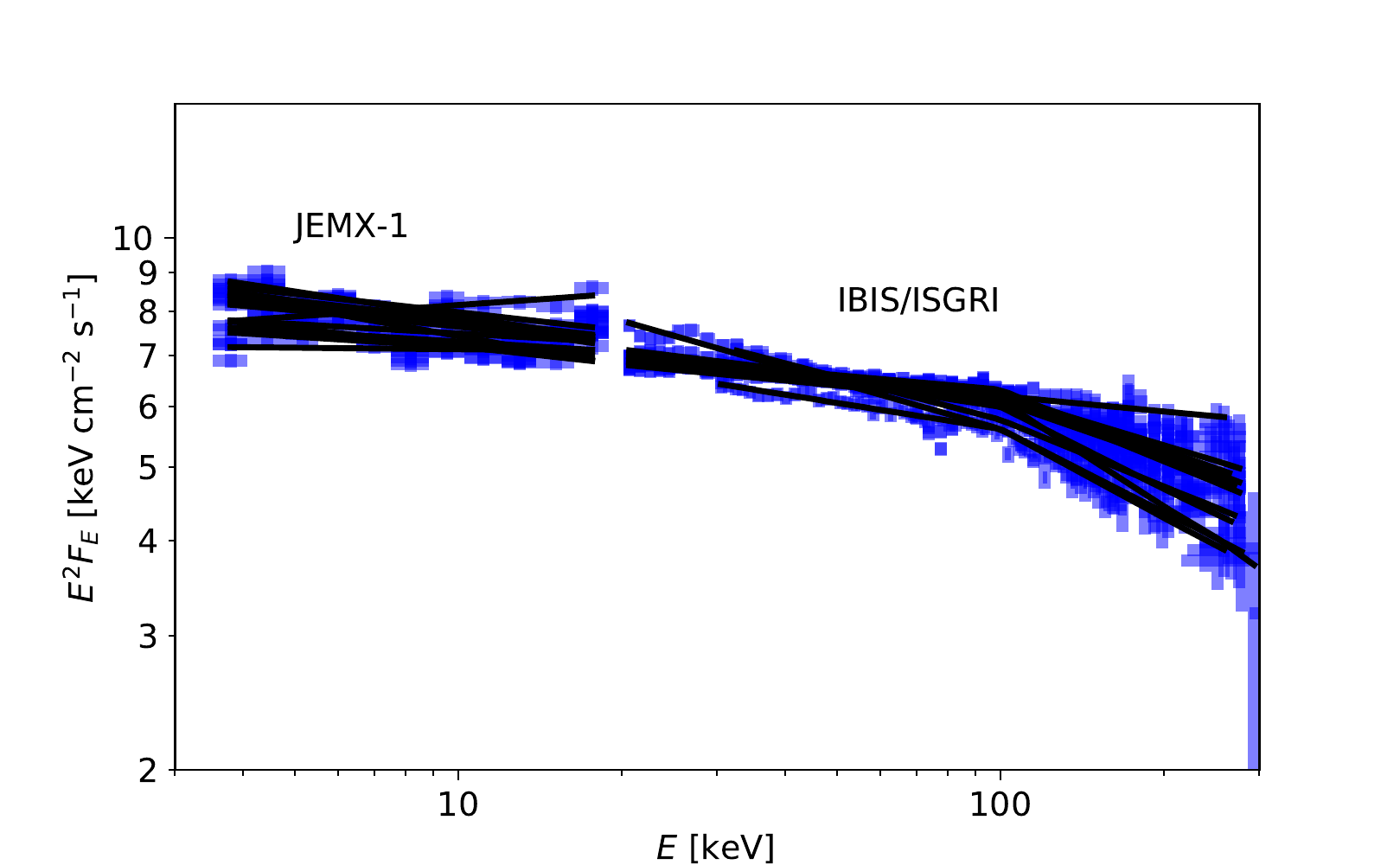}
    \caption{Display of the Crab unfolded spectra extracted during each year from 2003 to 2018
     from a samples of 50 randomly selected ScWs.
     The fitted model (in black) is a double broken power law with breaks fixed at 20 and 100\,keV.
     The analysis results could be re-generated through \href{https://github.com/oda-hub/oda_api_benchmark/tree/master/examples}{oda-hub} notebook \href{https://github.com/oda-hub/oda_api_benchmark/tree/master/examples/Fit-Crab-Spectra.ipynb}{Fit-Crab-Spectra.ipynb} run after \href{https://github.com/oda-hub/oda_api_benchmark/tree/master/examples/Crab_spectra_lc.ipynb}{Crab\_spectra\_lc.ipynb}.
    }
    \label{fig:Crab_spectrum}
\end{figure}

The ODA interface allows us to extract also images, spectra and lightcurves of JEM-X instrument, using the same set of parameters as for \isgri.
Fig.~\ref{fig:crab_image_jemx} shows the 3-20 keV image of Crab extracted from 50 randomly selected ScWs with pointing directions within $5^\circ$ from Crab (to
assure that the source is in the field-of-view of JEM-X.

Crab is detected with significance in excess of $10^3\sigma$ in the image. Two sources are detected with significance larger than 20 in the image: Crab and an
X-ray binary 1A 0535+262. We use a catalog containing these two sources for the spectral extraction.

\begin{figure}
    \centering
    \includegraphics[width=\linewidth]{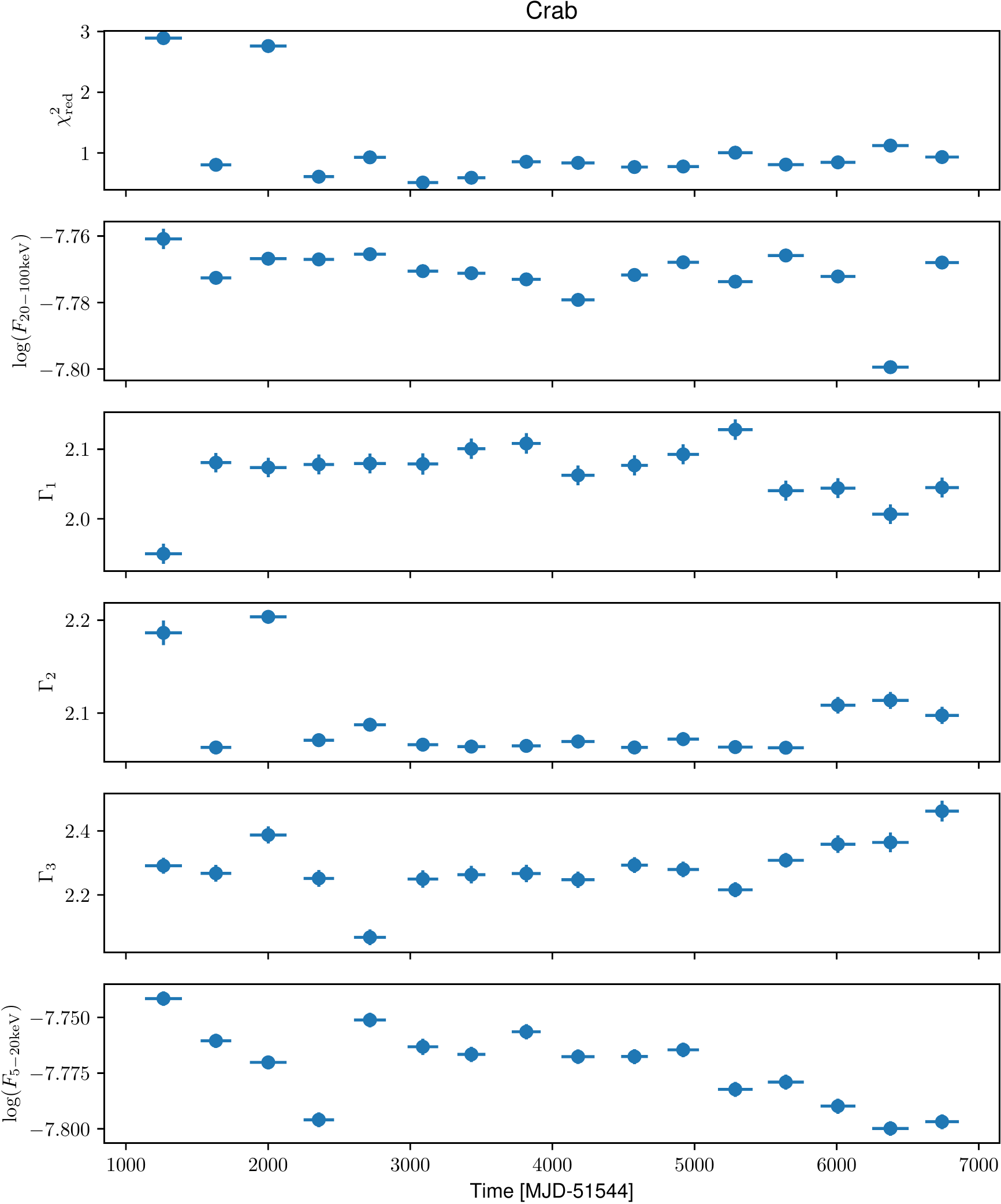}
    \caption{Best-fit spectral parameters with uncertainty intervals at 68\% confidence level on a single parameter and the reduced $\chi^2$ fit statistics of
        the Crab pulsar and nebula \isgri $+$ JEM-X1 spectra averaged on a year-by-year basis; the first point refers to 2003 and the last one to 2018. Data range is from 3.5 to 20 keV for JEM-X and 20--300 keV for \isgri. The
        model is a broken power law with break energies fixed at 20 and 100 keV; $\Gamma_{1-3}$ are the spectral indexed at increasing energies. We allowed
        spectra to assume different normalization through the flux
        parameter that we report for 5--20 keV (JEM-X1) and 20-100 keV (\isgri).
        The resulting number of degrees of freedom is 55 before 2016 and 112 after 2016.}
    \label{fig:crab_spec_results}
\end{figure}

Fig.~\ref{fig:Crab_spectrum} shows combined \isgri $+$ JEM-X1 unfolded spectra of the spectrum of Crab, extracted from the 50~ScW data sets on year-by-year basis
from 2003 to 2018. \change{We have modeled the spectra with a broken power law with two break energies fixed at 20 and 100\,keV.
The former value is meant to catch possible differences between JEM-X1 and \isgri spectral responses; the latter is a reasonable approximation for the
most probable spectral shape of Crab \citep{Jourdain2009}.
We applied a systematic factor of 2\%
to all spectral data, limited JEM-X1 data
between 5 and 20 keV, \isgri data
between 20 and 300 keV. The number of degrees of freedom is 55 for OSA10.2 (before 2016) and
112 after 2016 (OSA11.0).
Fig. \ref{fig:crab_spec_results} shows the year-by-year change of the Crab spectrum measurements:
our results are roughly in-line with previous findings \citep[e.g.,][]{crab}.
Formally non-acceptable fits are present in 2003 and 2005 with an anomalous power-law spectral index $\Gamma_2$ (20--100 keV), due to a
change of low threshold in the \isgri detector, which is currently taken into account in a sub-optimal way by the calibration files. Similarly, an anomalous
low flux in 2017 for \isgri is due to calibration issues being currently investigated by part of our team in a parallel effort.}

\subsection{Extremely bright source: V404 Cygni}

\begin{figure}
\includegraphics[width=\columnwidth]{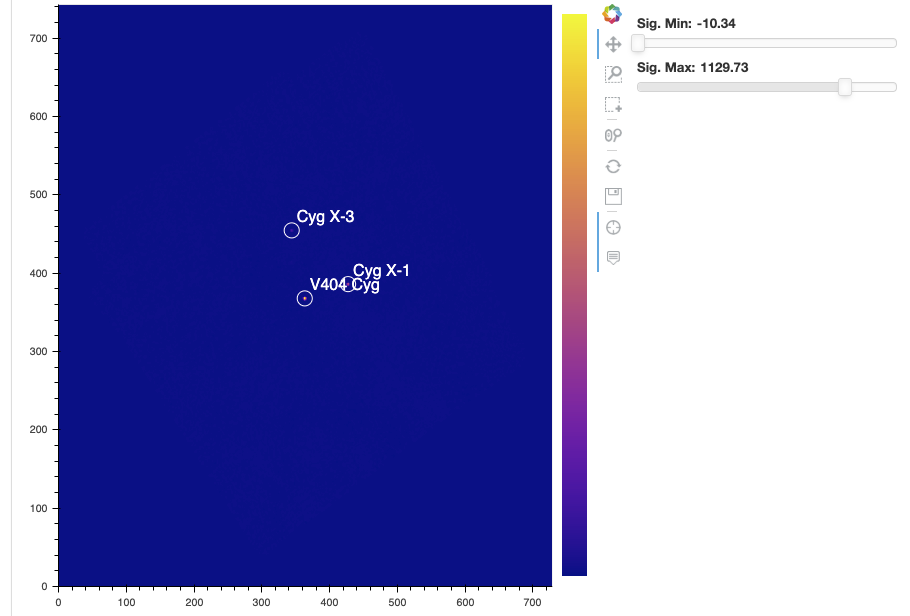}
\caption{Significance map of V404 Cygni region in 20-40 keV band. The image can be re-generated via URL \url{https://doi.org/10.5281/zenodo.3634669}. }
\label{fig:v404_image}
\end{figure}

V404 Cygni is a microquasar {that} underwent a spectacular outburst in 2015 during which the source flux has reached 50 Crab  \citep{v404,v404_rodriguez}. Such high flux might pose challenges for data analysis because of saturation {or} pileup effects {that} need to be properly taken into account.

To validate the performance of ODA for the bright source case, we have reproduced the results of the analysis of V404 Cyg reported by \citet{v404_rodriguez}.

At the first stage we assess the overall evolution of the source throughout the activity period {that} lasted from 2015-06-20T15:50:00.0 UTC to 2015-06-25T04:05:59.0 UTC, analyzed by \citet{v404_rodriguez}. Entering this time interval into the ODA interface via the \href{\odaurl ?src_name=V404 Cygni&RA=306.015917&DEC=33.867194&T1=2015-06-20T15:50:00.0&T2=2015-06-25T04:05:59.0&T_format=isot&instrument=isgri&osa_version=OSA10.2&radius=15&use_scws=no&E1_keV=20&E2_keV=40&query_type=Real&detection_threshold=30&product_type=isgri_image}{the URL}, we let the system select randomly 50 ScWs for the analysis with pointing directions within 10 degrees from the source direction and produce a mosaic image shown in Fig.~\ref{fig:v404_image}. The source is the brightest one in the image detected with significance exceeding 1000$\sigma$. The next brightest source in the field is Cyg X-1. A very strong source produces ``ghost'' images due to the specific of the coded mask imaging technique. This could be readily seen {in the} example of V404 Cyg by extracting all the sources in the image detectable at significance threshold above 5$\sigma$: this would result in very large amount of ``ghosts'' {that} would appear in the resulting catalog as ``NEW'' sources.

\begin{figure*}
\includegraphics[width=\linewidth]{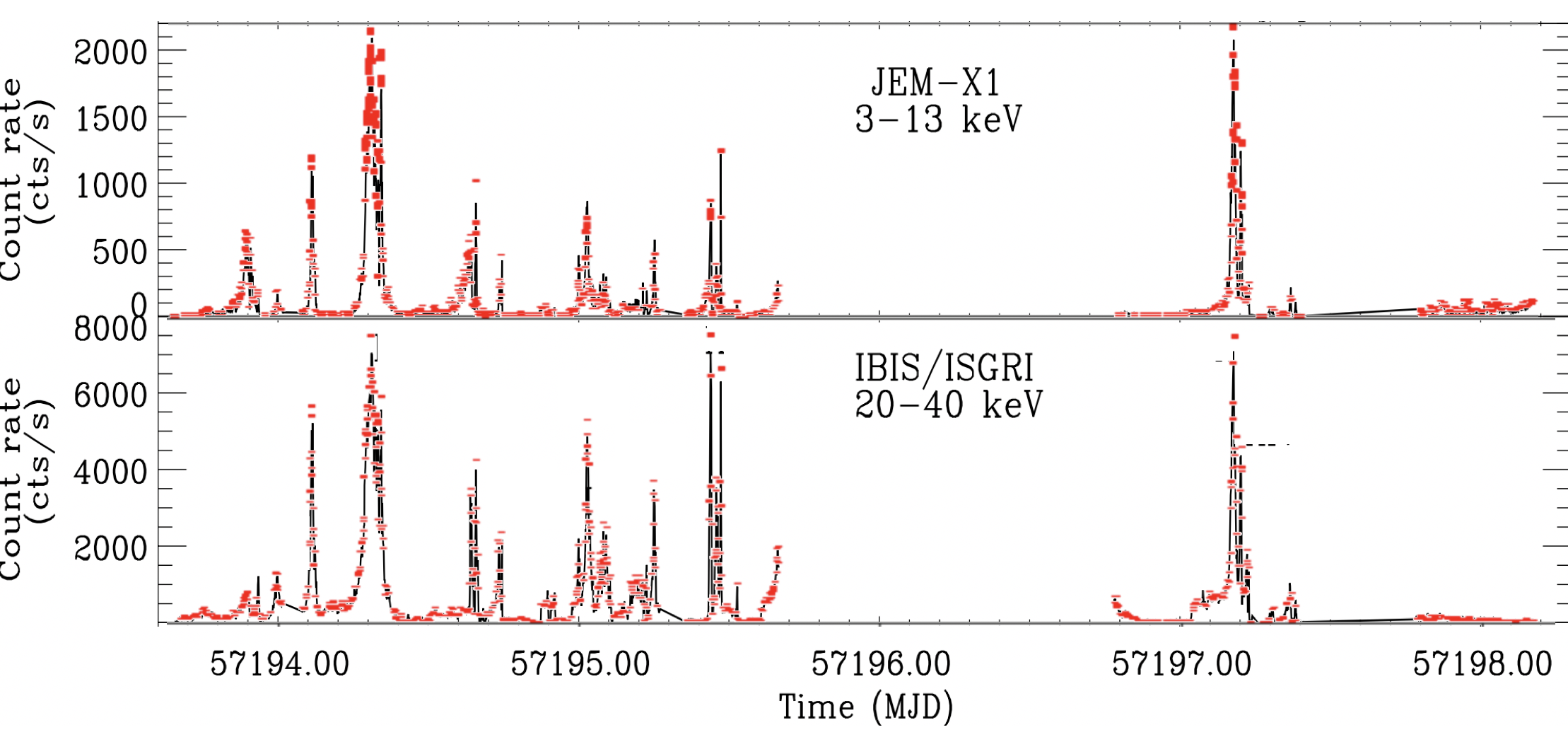}
\caption{JEM-X1 (top) and \isgri (bottom) Lightcurves of V404 Cygni during 2015 flaring period (red data points). Black lines are from Ref. \citep{v404_rodriguez}.}
\label{fig:v404_lc}
\end{figure*}

Fig. \ref{fig:v404_lc} shows the JEM-X1 and \isgri lightcurves of V 404 Cyg extracted for this period. Given the problem of detection of ``ghost'' sources around a very bright source, it is important to correctly define an input source catalog for the {light curve} production. Otherwise, the catalog {that} would be produced based on the imaging step of {the} analysis would include all the ``ghost'' sources in the procedure of fitting the detector image{,} which would lead to either wrong flux calculation{s} or larger error estimate{s} compared to the flux measurement{s} {including} only real sources.

From {F}ig. \ref{fig:v404_lc}, one {can} notice that the source underwent several short time flares. The overall evolution of the source flux inferred from ODA is practically indistinguishable from that calculated by  with that reported by \citet{v404_rodriguez}. This is clear from direct comparison of the two results, shown in Fig. \ref{fig:v404_lc}, where black color shows the lightcurve of \citet{v404_rodriguez} and red color shows the result extracted using ODA.

\subsection{Crowded field: GX~5$-$1}

\begin{figure}
\includegraphics[width=\columnwidth]{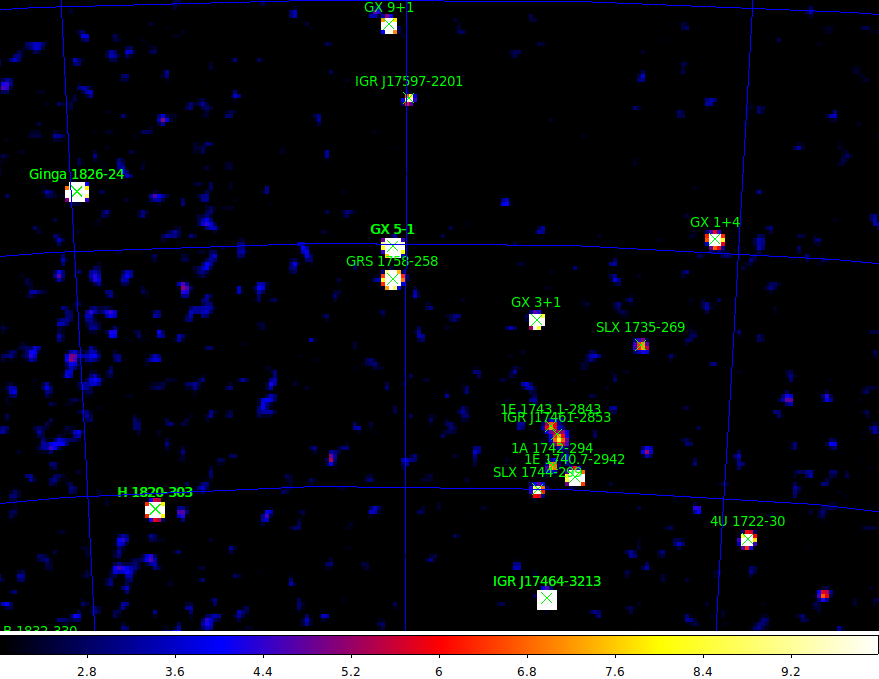}
\caption{Zoom of the significance map of the sky region around GX 5$-$1 in the 20--30\,keV energy band obtained from a selection of 44 science windows belonging to the GPS and GCDE programs carried out in 2013 with a poiting offset of less than three degrees from GX~5$-$1. This image can be obtained from the following \href{https://www.astro.unige.ch/cdci/astrooda_?DEC=-41.593417&E1_keV=20&E2_keV=30&RA=257.815417&T1=2003-03-15T23:27:40.0&T2=2003-03-16T00:03:15.0&T_format=isot&detection_threshold=7.0&instrument=isgri&osa_version=OSA10.2&product_type=isgri_image&query_status=ready&query_type=Real&radius=15&scw_list=005600620010.001,005600630010.001,005600800010.001,005600810010.001,005900170010.001,006100710010.001,006100800010.001,006100860010.001,006100870010.001,006100880010.001,006100960010.001,006100970010.001,006101030010.001,006101040010.001,006200200010.001,006200700010.001,006200770010.001,006200780010.001,006200790010.001,006200850010.001,006200860010.001,006300230010.001,006300240010.001,011900910010.001,011900920010.001,011901070010.001,011901080010.001,012000600010.001,012000610010.001,012000700010.001,012000760010.001,012000770010.001,012000780010.001,012000860010.001,012000870010.001,012000930010.001,012000940010.001,012000950010.001,012001020010.001,012001030010.001,012001110010.001,012100110010.001,012100120010.001,012200110010.001&src_name=4U+1700-377&use_scws=form_list}{URL}.
}
\label{fig:GX5-1_image}
\end{figure}

To verify the performance of ODA in the crowded field, we consider an example of GX~5$-$1, a persistent bright low-mass X-ray binary with a neutron star. It is located in the inner Galaxy region in a ``crowded'' field with many bright sources. In such situation, OSA has to take into account simultaneously all the bright sources while modeling superposition of the shadows of different sources on the detector. If this is not done properly, the signal of the source of interest could be contaminated by the overlapping shadows of unaccounted sources. Moreover, if bright sources are not included in the sky model
for spectral or light curve extraction, their photons will be erroneously assigned to other sources or background, hampering a reliable estimate.
However, neglecting weak sources is a minor concern if one is interested in
bright sources, as contamination can reach at most a few percents of the contaminant's flux in the worst cases.

To provide a direct comparison with the published results by \citet{Paizis2005}, we reproduced their selection of 44 science windows when the source was observed by the ``GPS'' and ``GCDE'' programs in 2013 with a maximum offest of three degrees. We made first an image in the 20--30\,keV band (Fig.~\ref{fig:GX5-1_image}) to select sources with a minimum significance of 7 for the subsequent spectral extraction. This resulted in a catalog of 25 sources. It should be noted that, owing to OSA construction, some sources might appear multiple times in the catalog, so it is necessary to delete duplication
from the catalog used as input in the following steps.

We performed the spectral extraction of both JEM-X2 amd \isgri data using the python API with the same selection of science windows. The dead-time and vignetting corrected exposures for JEM-X2 and  \isgri are 40.6, and 57.2\,ks, respectively. We added 1\% flux systematic to \isgri and 3\% to JEM-X2, we ignored JEM-X2 data below 3 and above 20 keV; \isgri data below 20 and above 70 keV, owing tot he partucular spectral shape. We modelled the joint spectrum with the same spectral models of \citet{Paizis2005}: the ``western'' model made of a \texttt{compTT+bbodyrad} and the ``eastern'' model, made of \texttt{compBB+diskbb} using \textsc{Xspec}. We have also introduce a cross-normalization factor fixed to one for \isgri, to account for non-simultaneity of some data.

We determined the 1$\sigma$ confidence ranges of parameters using a Monte Carlo Markov Chain with the Goodman-Weare  algorithm  \citep{Goodman2010} \change{as implemented in Xspec v. 12.11.0, and taking the 16, 50, 84\% percentiles of the posteriors as lower, central, and upper values. For the chain, we used 40 walkers, a length of 26\,000 and a burn-in phase length of 6000.} Results are presented in Table~\ref{tab:gx51}: both model represent well the spectra. However, the evolution of instrument calibration and analysis algorithms has lead to significant differences in the parameter values between the 2005 work, made with OSA v4.2 and ours, performed with OSA v.10.2. This is inherent to OSA and not directly related to the interface used to extract these spectra. A python notebook with the full workflow is available at \href{https://gitlab.astro.unige.ch/reproducible_INTEGRAL_analyses/gx5-1}{this URL}.

\begin{table}
	\renewcommand{\arraystretch}{1.3}
	\centering
	\caption{Best-fit parameters of spectral modelling of GX 5$-$1 in analogy with \citet{Paizis2005}.}
\begin{tabular}{lr@{}lc}
	\hline
	\hline
\multicolumn{4}{c}{Western model}	\\
\hline
$kT_\mathrm{BB}$ & 2.05 &$\pm$ 0.10 & keV \\
norm$_\mathrm{BB}$ & 26&$_{-8}^{+12}$ & $(\mathrm{km/d_{10}})^2$ \\
$T_0$ & 1.06 &$\pm$ 0.03 & keV \\
$kT_e$ & 5.4 &$\pm$ 0.5 & keV \\
$\tau$ & 1.5&$_{-0.2}^{+0.3}$ &  \\
norm & 1.1 &$\pm$ 0.1  & \\
factor & 0.76 &$\pm$ 0.03 & \\
$\chi^2$/d.o.f. &  18&/19 & \\
	\hline
\hline
\multicolumn{4}{c}{Eastern model}	\\
\hline
$T_\mathrm{in}$ & 1.99 &$\pm0.04$ & keV \\
norm$_\mathrm{disk}$ & 131&$_{-10}^{+11}$ &  $(\mathrm{km/d_{10}})^2$  \\
$kT$ & 3.0 &$_{-0.1}^{+0.2}$ & keV  \\
$\tau$ & 0.27&$_{-0.12}^{+0.08}$ & \\
norm & 4.1&$_{-1.5}^{+2.0}$ &  \\
factor & 0.74&$_{-0.04}^{+0.03}$ & \\
$\chi^2$/d.o.f. & 19&/20 & \\
\hline
Flux (5--20\,keV)\tablefootmark{a} & 1.28&$\pm0.02\times 10^{-8}$ & $\mathrm{erg\,s^{-1}\,cm^{-2}}$ \\
Flux (20--100\,keV)\tablefootmark{a} & 3.97&$\pm0.06\times 10^{-10}$ & $\mathrm{erg\,s^{-1}\,cm^{-2}}$ \\
\hline
\end{tabular}
\tablefoot{
	\tablefoottext{a}{Fluxes from both models are compatible within the uncertainties, so we report them only once.}
}

\label{tab:gx51}
\end{table}

\begin{figure}
\includegraphics[width=\columnwidth]{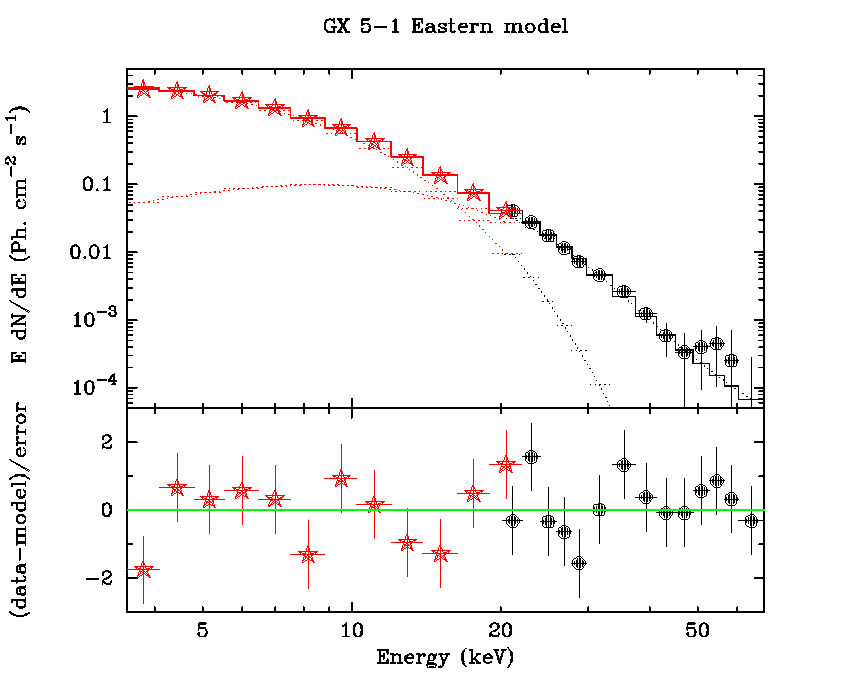}
\caption{Time-averaged spectrum of GX 5--1 extracted from the same 44 science windows as \citet{Paizis2005}, collected in 2003 with a maximum pointing offset of 3$^\circ$ from the source. Red stars and black circles represent JEM-X2 and \isgri data, respectively.}
\label{fig:GX51_spectrum}
\end{figure}

\subsection{3C 279}
\label{sec:3c}

In spite of its large bolometric luminosity and powerful jet \citep{3C279_1}, the active galactic nucleus (AGN) 3C 279 is a weak source for \integral. It has
hard spectrum in the hard X-ray energy band. Its flux is at the level of {the} sensitivity limit of \isgri and its detectability depends on the source state.
One of the flaring episodes occurred in 2015 and {\integral} observations {of 3C 279} during this episode were {obtained as a} Target-of-{O}pportunity campaign.
The results of data analysis for this TOO are described by \citet{3c279}.

Following the same approach as for other sources, we first performed an exploration of the source behavior throughout the 15 year time span of the data.
However, this source would likely be only marginally detected in any 50 ScW exposure, and no assessment of the variability pattern is possible in this way.

Instead, the full dataset has to be explored to find the periods during which the sources is detected. We use the API access to ODA to work with the datasets
longer than 50 ScWs in the follwoing way. As for the Crab lightcurve case in Sect.~\ref{sec:crab}, the Jupyter notebooks for the 3C 279 analysis {can} be
launched from the \href{https://github.com/oda-hub/oda_api_benchmark}{oda-hub/oda\_api\_benchmark}

GitHub repository, which is integrated with {the} Binder interactive notebook service\footnote{\href{https://mybinder.org}{https://mybinder.org}}.
\href{https://mybinder.org/v2/gh/oda-hub/oda_api_benchmark/master}{Launching the binder} using the ``launch binder'' button in the
\href{https://github.com/oda-hub/oda_api_benchmark}{oda-hub/oda\_api\_benchmark} repository and choosing the notebooks for 3C 279 lightcurve and spectra found
in {\tt examples} {makes it} possible to generate the results described below online.

\begin{figure}
\includegraphics[width=\linewidth]{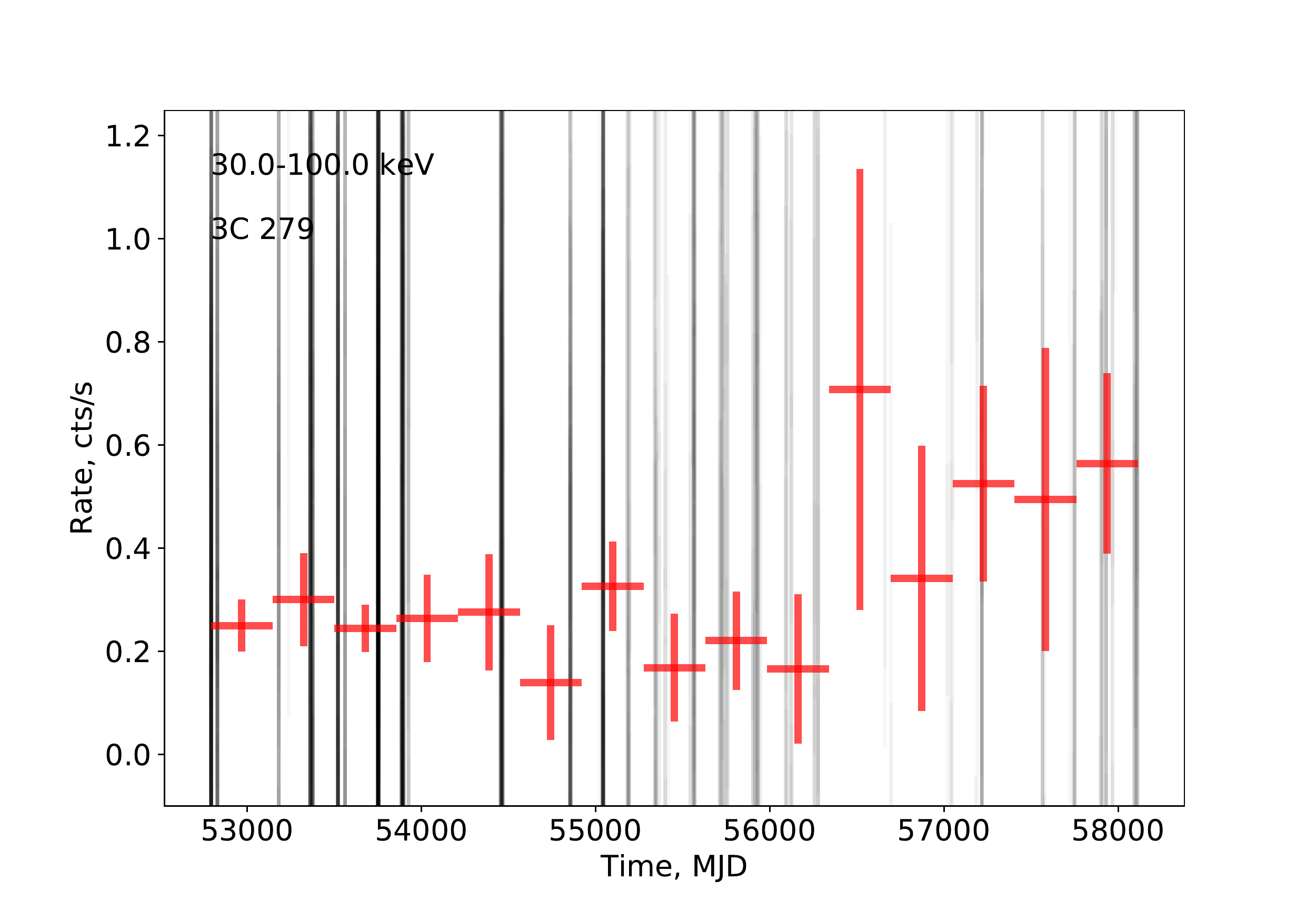}
\caption{Lightcurve of 3C 279 on 15 yr time span. Grey vertical lines show exposure periods of the source. The notebook \href{https://github.com/oda-hub/oda_api_benchmark/blob/master/examples/3C279_lc.ipynb}{3C279\_lc.ipynb} for the calculation of the lightcurve could be launched using this \href{https://mybinder.org/v2/gh/oda-hub/oda_api_benchmark/master}{URL}.}
\label{fig:3c279_lc1}
\end{figure}

At the first stage of analysis, we determine the bright sources in the source field. We generate a mosaic image of the field and use the output catalog of the mosaic analysis, adding explicitly 3C 279 to the catalog, as an input step for the timing and spectral analysis. Using the resulting source catalog, we process all sets of 50 ScWs in sequence to obtain a long-term lightcurve of the source shown in Fig. \ref{fig:3c279_lc1}. From this figure one could see that the source is systematically detected throughout the entire 15 year time span. It shows moderate (if any) variability from year to year.

\begin{figure}
\includegraphics[width=\linewidth]{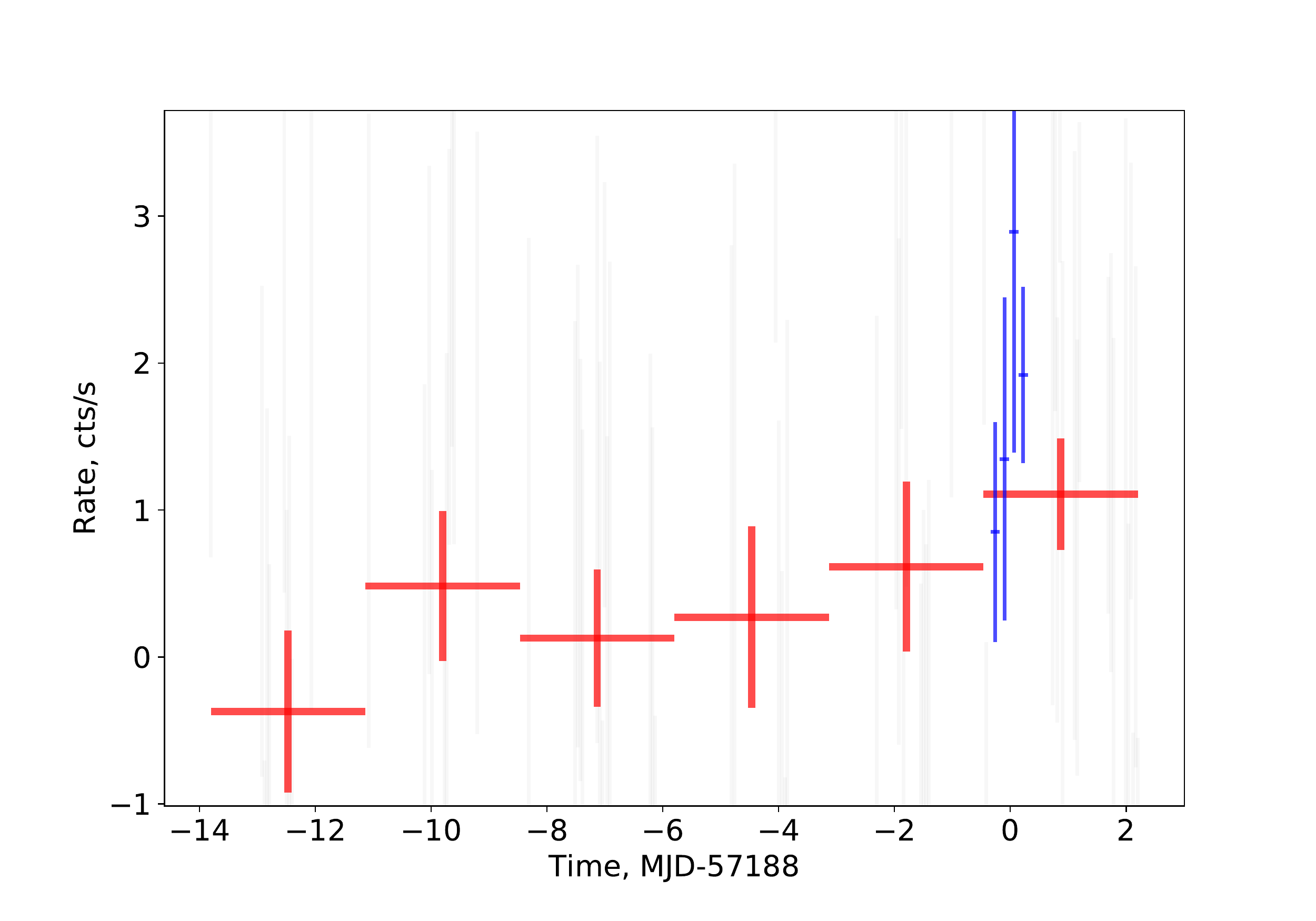}
\caption{Lightcurve of 3C 279 in 20-100 keV range for the flare period reported by \citet{3c279} (red points), extracted using the ODA API, and the \isgri lightcurve reported in fig. 1 of \citet{3c279} (blue points), corresponding to their \isgri data time span. . The notebook \href{https://github.com/oda-hub/oda_api_benchmark/blob/master/examples/3C279_lc_flare.ipynb}{3C279\_lc\_flare.ipynb} for re-generation of the result {can} be executed online at this \href{https://mybinder.org/v2/gh/oda-hub/oda_api_benchmark/master}{URL}{,} and is re-executable via Binder integration of oda-hub project.}
\label{fig:3c279_lc}
\end{figure}

The 2015 flare of the source reported by \citet{3c279} is identifiable as the highest flux point in the lightcurve in Fig. \ref{fig:3c279_lc1}. More detailed view of the lightcurve for the flaring episode discussed by \citet{3c279} is shown in Fig.~\ref{fig:3c279_lc}, where we plot (red points) the ODA API lightcurve, extracted for the full range of the flaring period investigated in \citet{3c279}, and the \isgri light curve  reported in Fig.~1 of \citet{3c279}. The average flux of the \citet{3c279} lightucurve is compatible with our bin overlapping the same time span
The average count rate is at the level of 1 ct/s{,} which agrees with the published value. This lightcurve {can} be re-generated using the same lightcurve extraction notebook as for the long-term lightcurve of the source, changing the time interval to focus on the flaring period{,} July 2015{,} and adjusting the energy range. This shows how the notebook available for on-the-fly re-deployment via the \href{https://github.com/oda-hub/oda_api_benchmark}{oda-hub/oda\_api\_benchmark} web page {can} be re-used for refinement or re-use of the analysis for different energy range{s} or different source{s}.

\begin{figure}
\includegraphics[width=\linewidth]{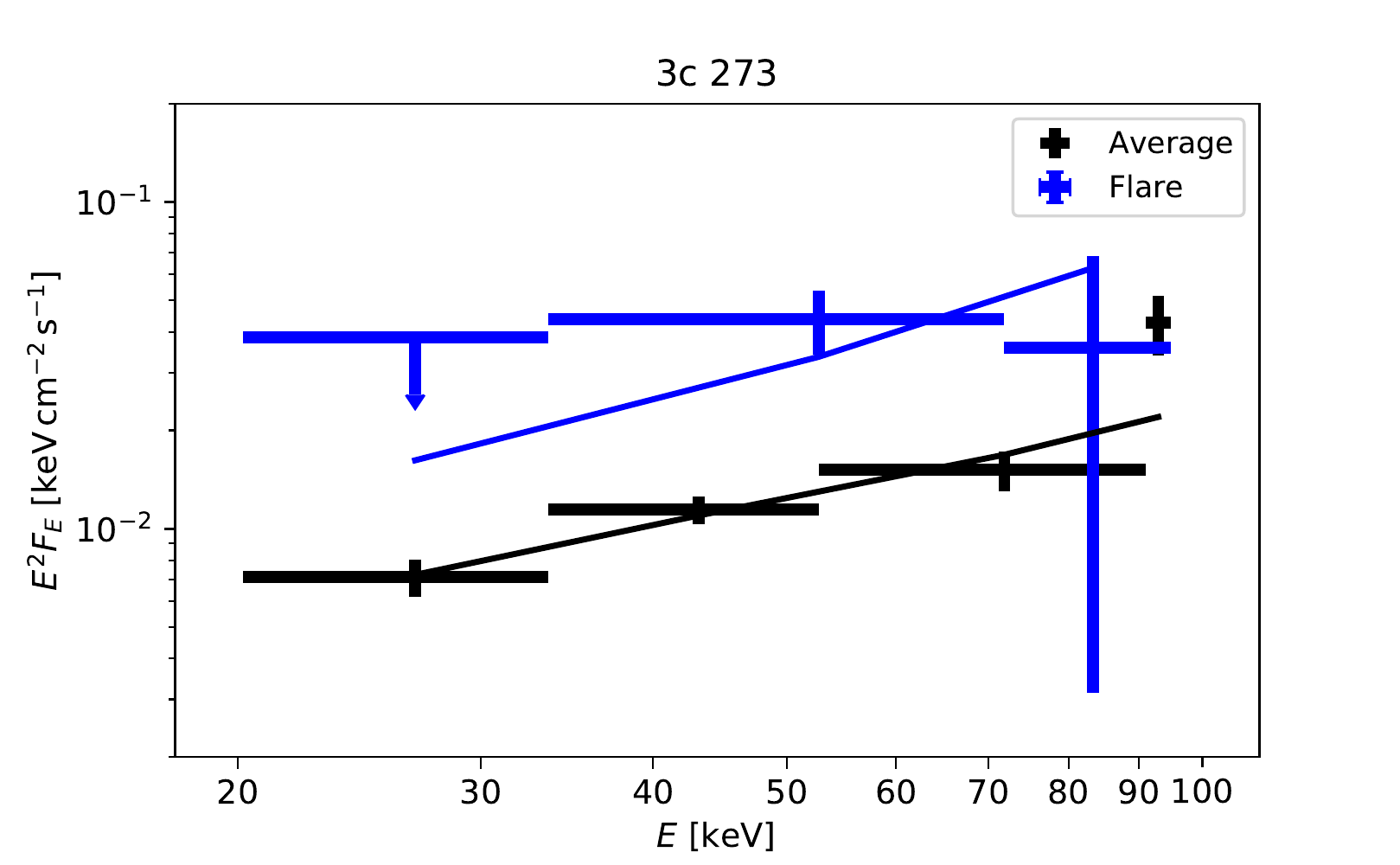}
\caption{Comparison of time-averaged spectrum of 3C\,279 (black) with the spectrum of the flaring state observed during the  TOO period reported by
    \citet{3c279} in blue; the upper limit is at 3$\sigma$ confidence level. The respective models are represented as solid lines.
    The notebook \href{https://github.com/oda-hub/oda_api_benchmark/blob/master/examples/3C279_spectrum.ipynb}{3C279\_spectrum.ipynb} for the calculation of the spectra {can} be launched using this \href{https://mybinder.org/v2/gh/oda-hub/oda_api_benchmark/master}{URL}.}
\label{fig:3c279_spectrum}
\end{figure}

Fig. \ref{fig:3c279_spectrum} shows a comparison of the time-averaged spectrum of the source with the flaring state spectrum. We have used the same spectral range extraction of 20-100 keV, and the same spectral model (flux-pegged power-
law model \texttt{pegpwrlw}\footnote{Based on Xspec package fitting (\url{https://heasarc.gsfc.nasa.gov/xanadu/xspec/manual/node207.html}).}) as in  \citet{3c279}.
The spectral fit of the flaring state spectrum shown in Fig. \ref{fig:3c279_spectrum} has {a} very hard slope  $\Gamma=0.81_{-0.11}^{+1.71}$ and a flux, in the 18-55 keV band, of $4_{-3}^{+5}\times 10^{-11}\,\mathrm{erg\,cm^{-2}s^{-1}}$ \change{( both reported at 90\% c.l.)}, which are consistent with \citet{3c279}, where the authors report a photon index of $\Gamma=1.08_{-0.15}^{+1.98}$  and a flux of
$8_{-6}^{+10}\times 10^{-11}\,\mathrm{erg\,cm^{-2}s^{-1}}$.

\subsection{NGC 2110}

NGC 2110 is an example of moderately bright Seyfert galaxy, i.e. a representative of typical hard X-ray bright and persistent AGN  \citep{agn}.
AGN of this type are the most abundant extragalactic sources observed by \integral.

\begin{figure}
\includegraphics[width=\linewidth]{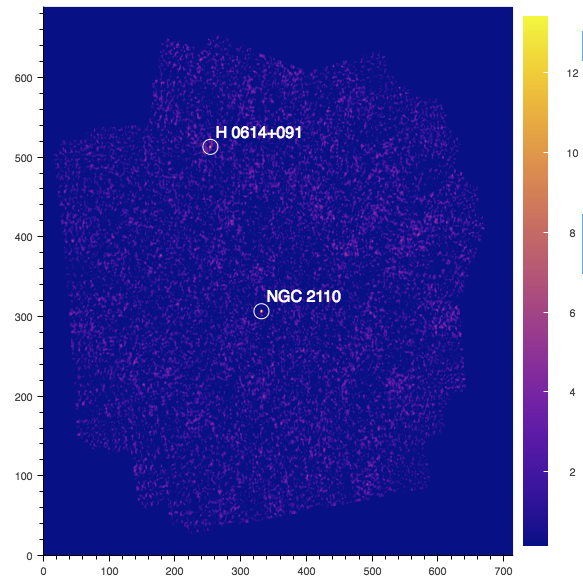}
\caption{20-40 keV image of {the} NGC 2110 region extracted from a set of 50 randomly selected ScWs. The image could be regenerated via the URL
\url{https://doi.org/10.5281/zenodo.3634584}.
}
\label{fig:NGC2110_image}
\end{figure}

Fig. \ref{fig:NGC2110_image} shows the significance image of the source region extracted from a set of 50 random ScWs spread over 15 years of \integral
operations. One {can} see that NGC 2110 is the brightest source in the field. Its detection significance reaches $\simeq 17\sigma$ in the exposure $T\simeq
90$~ks. The dimmer source H 0614+091 is detected at significance level {$10\sigma$}, {and} there {is} no other source in the field  detected with significance
exceeding $5\sigma$.

\begin{figure}
\includegraphics[width=\linewidth]{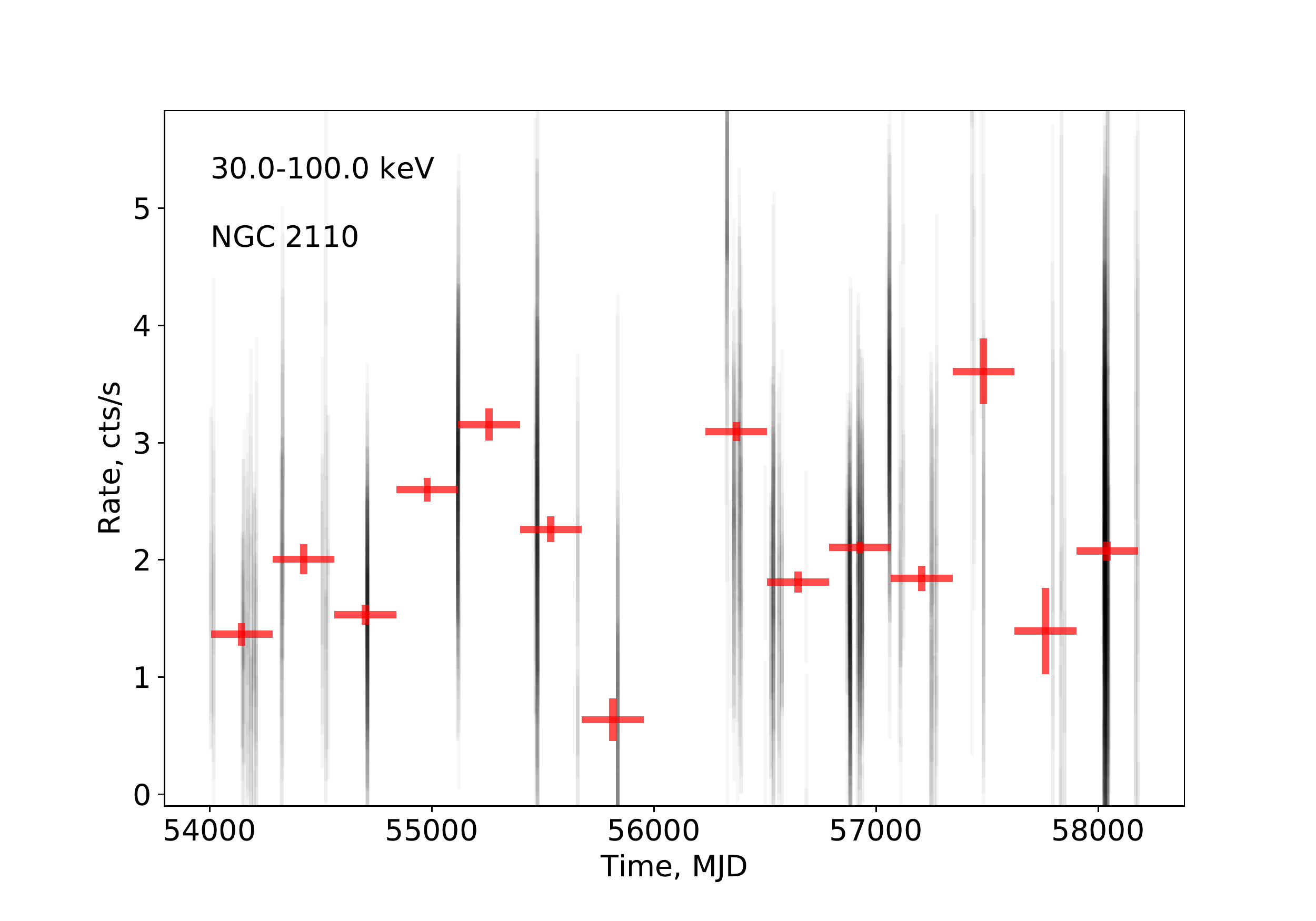}
\caption{Rebinned lightcurve of NGC 2110 extracted from the sequence of 50 ScW-long intervals using the notebook \href{https://github.com/oda-hub/oda_api_benchmark/blob/master/examples/lc_longterm.ipynb}{lc\_longterm.ipynb}. The result could be  re-generated via notebook deployment on Binder at this  \href{https://mybinder.org/v2/gh/oda-hub/oda_api_benchmark/master}{URL}. }
\label{fig:NGC2110_lc}
\end{figure}

Fig. \ref{fig:NGC2110_lc} shows the long-term lightcurve of the source extracted from the sequence of 50 ScW datasets pointed within 10 degrees from the source,
using the same notebook as for the 3C 279 analysis, part of the
\href{https://github.com/oda-hub/oda_api_benchmark/blob/master/examples/lc_longterm.ipynb}{the oda-hub/oda\_api\_benchmark} repository.
The statistical uncertainty of the flux measurement in individual ScWs is too large and it is not possible to follow the long-term evolution of the source
spectrum with ScW-by-Scw binning. Taking this into account, we have  rebinned the lightcurve into wider time bins. Several variability episodes are clearly
identifiable with such binning in the long-term lightcurve.

\begin{figure}
\includegraphics[width=\linewidth]{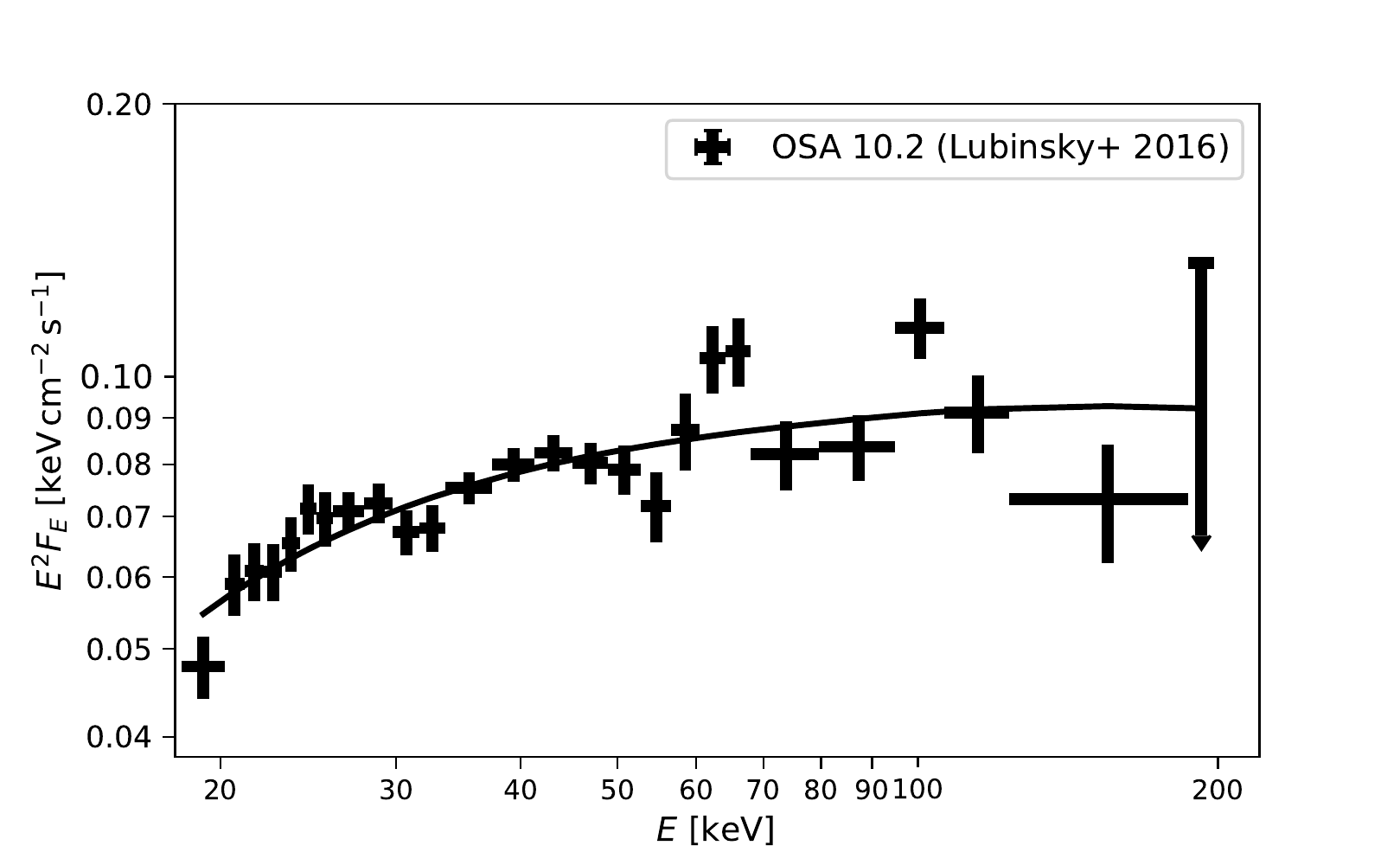}
\includegraphics[width=\linewidth]{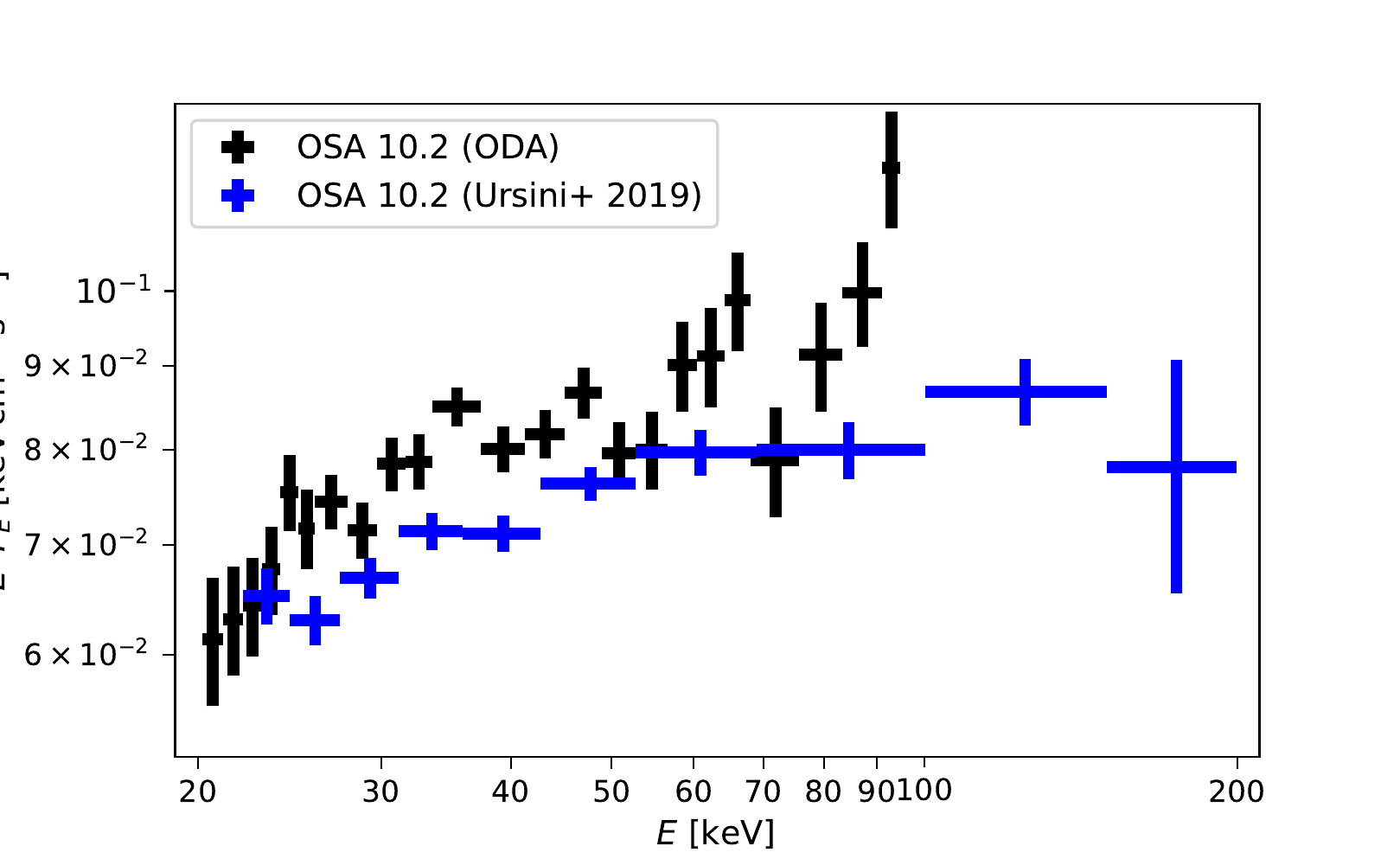}
\caption{Top: Spectrum of NGC 2110 extracted from the stacking of spectra obtained for 50 ScW sets in the time periods used in \citep{ngc2110},
    using OSA 10.2. Bottom: Comparison of the
    spectra of NGC 2110 extracted from the stacking of spectra obtained for 50 ScW sets  in the time periods used in \citep{Ursini_2019} and the spectra
    published in \citep{Ursini_2019}.
The spectra {can} be re-generated online using the notebook
\href{https://github.com/oda-hub/oda_api_benchmark/blob/master/examples/spectrum_longterm.ipynb}{spectrum\_longterm.ipynb} and
\href{https://github.com/oda-hub/oda_api_benchmark/blob/master/examples/spectrum_longterm_Ursini19.ipynb}{spectrum\_longterm\_Ursini19.ipynb}.
}
\label{fig:NGC2110_Lubinsky_2016_spectrum}
\end{figure}

We used the same approach as for 3C 279 for the extraction of the source spectrum with long exposure time. We extracted the spectrum of NGC 2110 stacking the
spectra in sequences  of 50 ScW long exposures and calculating the weighted average using the same notebook as for the extraction of the 3C 279 spectrum,
available and executable at the \href{https://github.com/oda-hub/oda_api_benchmark/blob/master/examples/spectrum_longterm.ipynb}{oda-hub/oda\_api\_benchmark}
project page. The resulting source spectrum with a total exposure of 2.4 Ms is shown in Fig.~\ref{fig:NGC2110_Lubinsky_2016_spectrum}. The figure shows two
separate spectra for the periods of applicability of OSA 10.2 and OSA 11.0. The physical origin of the hard X-ray emission from Seyfert galaxies is thought to
be due to the inverse Compton scattering of emission from an accretion disk in a hot corona characterized by certain temperature \citep{ngc2110}{.} Suppression
of the emission above 100 keV expected in this model is seen in the longer (1.8 Ms) exposure spectrum for OSA 10.2 applicability period. However,  the exposure
of OSA 11.0 period (0.6 Ms) is not sufficiently long
to constrain a high-energy cut-off in the spectrum.
In order to compare our results with those reported in \citep{ngc2110}, we have extracted ScWs for the same time span and \change{angular extraction cone
radius}. We have a final exposure of $\simeq 1714$ ks. \change{We stress that full reproduction of the analysis reported in  \cite{ngc2110}  is beyond the scope
of our work, because we are not using soft X-ray data to constrain the parameters of the \change{ \texttt{COMPPS} Xspec model, used to describe inverse Compton
emission.}
In our analysis we have used the same model as in \citet{ngc2110}, and we have fixed all the parameters to those reported in their analysis, except for
the electron temperature $kT_e$, the $y$-Compton parameter and the normalization. In detail, we have fixed the seed photons temperature $T_{bb}$ to a value of
10 eV, the amount of reflection $R$ to 0.63, and the geometry to the spherical case.
We have estimated  the $90\%$ parameters confidence range using  the Xspec (v. 12.11.0) MCMC implementation,  based on the Goodman-Weare algorithm
\citep{Goodman2010}, with 20 walkers, and a Cauchy proposal distribution, running a 10000 steps chain, with a burn-in phase of 3000 steps. We have run the
chain with
an initial state starting from the Xspec best fit solution.
We report the 0.05, 0.5, and 0.95 quantiles of the posterior distribution as lower, central, and upper values.
We find a temperature  $kT_e=3.5^{+1.3}_{-2.4}\times 10^2$ keV, a normalization constant $N=2.6^{+1.2}_{-1.0}\times 10^{8}$, and a value of the $y$-Compton
parameter of $y=1.10^{+0.19}_{-0.25}$,
compared to the values reported in  \cite{ngc2110}, of  $kT_e=2.3\pm0.5\times10^2$ keV, $N=1.8^{+0.4}_{-0.2}\times 10^{8}$, and $y=0.94^{+0.10}_{-0.09}$,
respectively}.
\change{ The spectrum  and the frequentist best fit model are reported in the top panel of Fig.~\ref{fig:NGC2110_Lubinsky_2016_spectrum}. }
As further benchmark we have compared the results from ODA to those published in \cite{Ursini_2019}. In this case  \change{ the authors have distributed online
the \integral/\isgri spectra extracted with OSA} used for their publication. We have extracted {\color{blue} the \integral/\isgri  spectra with ODA} for the
same time span and spectral window as in \cite{Ursini_2019}. The results are shown in the lower panel of Fig. \ref{fig:NGC2110_Lubinsky_2016_spectrum}.

\section{Analysis limitations}\label{sec:limitations}

In the current ODA implementation, the dispatcher defines which scientific analysis workflows can be executed through the frontend, and enables two versions of
\integral analysis software: standard OSA10.2 and OSA11.0. OSA10.2 can only be used for the analysis of the data taken before 2016. OSA11.0 can only be used for
the \isgri data since 2016 (at the time of writing), while it can be used for JEM-X during the full mission lifetime. This will change, as soon as updated \isgri
calibration files are made available. It should be remarked that the introduction of new releases of OSA will be made available.

The maximum number of ScW allowed in a single request is set to 50 (corresponding to up to 50 CPU-hours of computing), to reduce load on the system through too
long jobs. This might change if the back-end is deployed in a different location or more computing resources are made available. Large requests will also be
available for authenticated users.
As of now, it is possible to overcome this limitation by looping over successive sets of 50 ScWs as it is demonstrated in the examples of analysis in this paper.

The \integral data analysis within ODA platform entirely relies on scientific data analysis with OSA, modifying only the mechanism with which  individual
components of OSA are invoked (principally to allow for preservation and re-use of intermediate analysis results, as explained above). It means that \integral
analysis within ODA shares the limitations of OSA, and any future changes in OSA will be available in ODA.

Detailed scientific validation of the long-term evolution of the \integral instruments and the current status of the data analysis software with OSA goes beyond
the scope of this paper.

\section{Comparison with other Online Analysis Platforms}
\label{sec:odavsworld}

Several other online data analysis platforms offer similar services, some of them - for \integral instruments.

HEASARC Hera exposes much of the HEASoft capabilities (with no relevant support for \integral) as a web service. This  original and promising
development  has limitations of the service resources.
Hera was, in part, an inspiration for the most superficial scheme of  \integral ODA back-end.

\swift/XRT online data analysis\footnote{\url{https://www.swift.ac.uk/user_objects/}} is a particularly successful example of an online analysis and has
encouraged adoption of this strategy for other instruments.  \xmm Newton's RISA\footnote{\url{https://www.cosmos.esa.int/web/xmm-newton/xsa}} is conceptually
similar to  \swift/XRT online analysis,
but is, arguably, less well known, perhaps due to  different strategy of observation scheduling in \xmm: while \swift is favoring a large number of short public
observations with quick impact,
\xmm observations are often part of larger campaigns and private observations, favoring traditional offline analysis.

Both \swift/XRT and \xmm  telescopes feature focusing mirrors, and their data analysis workflows require much smaller computing resources than \integral.
Coded mask telescopes feature much larger FoV than the focusing ones - which in turn results in a larger number sources to be considered by the decoding process.  In addition, their PSF is highly non-trivial and generally spread over the entire FoV.
The combination of the typically long observations and dependency on a model with potentially large number of sources means that online analysis of a coded mask telescopes is resource-hungry, and has to adopt a very different approach to handling data analysis workflows and pre-computed data.

\integral SPIDAI provides  online data analysis for \integral/SPI.
SPI data analysis requires considerable re-analysis for each new request. This is in principle similar to ODA \isgri and JEM-X workflows, but without the
comparably extensive re-use of pre-computed results. SPIDAI analysis can avoid this additional complexity since only relatively small number of sources are
sufficiently bright to be observed by SPI.

HEAVENS\footnote{\href{http://isdc.unige.ch/heavens/}{http://isdc.unige.ch/heavens/}}
follows very different approach from that of \integral ODA, SPIDAI, \swift/XRT, RISA, or Hera, and pre-computes the raw data using undisclosed customized
procedures, creating a space-energy-time map of celestial flux, with fixed pre-defined resolution in each dimension. This considerably speeds-up certain kinds
of analysis, but also implies that the scientific meaning of the HEAVENS output may be very different from that provided by expert-validated OSA. Furthermore, a
particular pre-defined selection of space-time-energy resolutions restricts certain kinds of analysis (e.g. light curves with time bins less than about
3000~seconds - the \integral pointing duration). Finally, pre-computing of the results is costly (both in computing time and human effort) and does not
necessarily privilege popular results, since it is not done in response to the user needs.
As a result - the pre-computed HEAVENS results are available with a large delay. Finally,
HEAVENS imposes additional severe restrictions on the output results, for example, the size of the requested image.

To the contrary, ODA \integral analysis runs on-demand OSA analysis based exclusively on publicly available tools.
Any  improvements  in OSA, officially validated by the instrument experts and the data center,
are immediately adopted by ODA (by providing an additional OSA version in the parameter selection).

Conversely, ODA operations expose some of the technical issues of OSA. While in the traditional offline analysis approach these issues would plague users with
poorly understood error messages, an equivalent message in ODA-wrapped OSA can be directed straight to the OSA software developers, and the patch will be made
available in the next OSA release (with a corresponding update to ODA).

\integral ODA provides a much larger set of results than that available in pre-computed databases (such as HEAVENS). For example, it is possible to produce
light-curves with small (down to 10~seconds) time bins and high-resolution spectra. More advanced users may upload custom source lists to use as sky model, to
provide the most reliable \integral results, as recommended by OSA cookbook. Pre-computing the results with this granularity would not be feasible in platforms
like HEAVENS.

The ODA platform is built as a cloud-native solution, designed to be adaptable to any modern computing environment. It is already open-source, and \change{can
be readily developed via addition of new components}. The platform consists of \change{a range of} independent components, following common  standards of
communication and interoperability. This means that the platform may grow beyond single deployment, for example \change{it can be deployed as a part of a
federated infrastructure} of  ESA DataLabs, \href{https://datalabs.esa.int}{https://datalabs.esa.int}.

ODA platform is strongly committed to interoperability and integration with  the European Open Science Cloud (EOSC)
\href{https://www.eosc-portal.eu}{https://www.eosc-portal.eu}, and has been validated as an EOSC service\footnote{\href{https://marketplace.eosc-portal.eu/services/astronomical-online-data-analysis-astrooda/}{https://marketplace.eosc-portal.eu/services/astronomical-online-data-analysis-astrooda/}}.

\section{Conclusions}

We have presented the new approach for the \integral data analysis which uses the cloud computing technology to enable deployment of data analysis workflows for
imaging, spectral and timing analysis online through a web interface in a web browser:
\href{https://www.astro.unige.ch/cdci/astrooda_}{https://www.astro.unige.ch/cdci/astrooda\_} or through a dedicated API {that can} be used in Python notebooks
executable locally or remotely from the \href{https://github.com/oda-hub/}{oda-hub} project Github repository. Such an approach provides an important boost for
reproducibility of results extracted from \integral data {and} possibilities of sharing and re-use of data analysis methods. Virtualisation of the data analysis
system also provides a viable solution for the long-term preservation of the data and analysis results.  This paper demonstrates  how reusable astronomical data
analysis workflows can be shared and embedded in publications.

Performance tests presented in this paper validate ODA for use in scientific publications making use of \integral data. ODA results are identical to those which
are obtained with OSA with parameters choices following standard recommendations of the OSA hand books.

\integral ODA  provides on-demand analysis of any \integral \isgri and JEM-X data, leveraging an almost complete set of OSA capabilities, yielding the results
identical to the expert-validated publicly released OSA. It is especially useful in the context of multi-wavelength and multi-messenger studies of variable
astronomical sources \citep{multimessenger}, because it provides read-to-use and flexible data products: images, spectra and lightcurves which can be adjusted
to specific details of source variability, observation periods of other instruments. It can also be used to explore long-term behaviour of multi-messenger
sources prior to their activity periods.

\bibliographystyle{aa}
\bibliography{001-references}
\end{document}